\documentclass[letterpaper,twocolumn,10pt]{article}
\usepackage{usenix-2020-09}

\usepackage{tikz}
\usepackage{amsmath}

\usepackage{filecontents}

\AtBeginDocument{%
  }

\newif\ifall    %
\newif\ifconf   %
\newif\ifsq     %
\newif\ifnonb   %
\newif\iftodos  
\newif\ifbld

\newif\ifsqCAP
\newif\ifsqVS
\newif\ifsqEN
\newif\ifsqTIT

\sqfalse
\sqCAPfalse
\sqENfalse
\sqVSfalse

\sqTITtrue

\usepackage{etex}
\usepackage{balance}
\usepackage{epstopdf}
\usepackage{placeins}

\usepackage{graphicx}
\usepackage{float}
\usepackage{dblfloatfix}
\usepackage{multirow}
\usepackage{rotating}
\usepackage[switch]{lineno} %
\usepackage{makecell}
\usepackage{tabulary}
\usepackage{parcolumns}
\usepackage{tikz}
\usetikzlibrary{tikzmark}

\usepackage{xpatch}
\expandafter\xpatchcmd
\csname pgfk@/tikz/every picture/.@cmd\endcsname
{\thepage}{\arabic{page}}{}{}

\tikzstyle{comment} = [draw, fill=blue!70, text=white, text width=3cm, minimum height=1cm, rounded corners, align=left, font=\scriptsize]
\tikzstyle{background_alg} = [draw, fill=blue!20, opacity=0.4, inner sep=4pt, rounded corners=2pt]

\usetikzlibrary{shapes}
\usetikzlibrary{plotmarks}
\usetikzlibrary{calc, fit}

\usepackage{enumitem}

\usepackage{amsmath,amssymb,amsfonts}
\usepackage{mathtools,mathrsfs}

\DeclarePairedDelimiter{\ceil}{\lceil}{\rceil}

\usepackage{mleftright}

\usepackage{soul}
\usepackage{fontawesome}
\usepackage{pifont}
\usepackage{textcomp}
\usepackage{booktabs}
\usepackage{url}
\usepackage{pbox}
\usepackage[normalem]{ulem}
\usepackage[10pt]{moresize}

\usepackage[capitalise]{cleveref}
\usepackage[utf8]{inputenc}
\crefname{section}{§}{§§}
\Crefname{section}{§}{§§}

\definecolor{aablack}{rgb}{0.4 0.4 0.4}

\ifsqCAP
\usepackage[font={bf, scriptsize}]{caption}
\usepackage[font={normalfont, scriptsize}]{subcaption}
\else
\usepackage[font={bf, footnotesize}]{caption}
\usepackage[font={normalfont, footnotesize}]{subcaption}
\fi

\newcommand{\vspaceSQ}[1]{\ifsqVS\vspace{#1}\fi}
\newcommand{\enlargeSQ}[1]{\ifsqEN\enlargethispage{\baselineskip}\fi}

\ifsqTIT
\usepackage[compact]{titlesec}
\titlespacing*{\section} {0pt}{1.75ex plus 1ex minus .4ex}{1.25ex plus .2ex}
\titlespacing*{\subsection} {0pt}{1.35ex plus 1ex minus .2ex}{0.75ex plus .2ex}
\titlespacing*{\subsubsection}{0pt}{1.35ex plus 1ex minus .2ex}{0.75ex plus .2ex}
\fi

\usepackage{xcolor}
\definecolor{darkgrey}{RGB}{70,70,70}
\definecolor{lightgrey}{RGB}{200,200,200}
\definecolor{lyellow}{RGB}{255,255,100}
\definecolor{llyellow}{RGB}{250,250,180}
\definecolor{lgreen}{RGB}{144,238,144}

\usepackage[customcolors]{hf-tikz}
\hfsetbordercolor{white}
\hfsetfillcolor{vlgray}

\definecolor{vlgray}{rgb}{0.77 0.77 0.77}
\definecolor{ablack}{rgb}{0.2 0.2 0.2}
\definecolor{vllgray}{rgb}{0.9 0.9 0.9}
\definecolor{bblue}{rgb}{0.7 0.7 0.99}

\usepackage{colortbl}

\usepackage{listings}

\ifsq
\else
\fi

\definecolor{hlL}{rgb}{0.8 0.8 0.99}

\newcounter{highlight}

\newcounter{hlLIR}

\newcounter{hlLIIR}

\newcounter{Ahighlight}

\usepackage{scalerel,stackengine}
\stackMath
\newcommand\rwh[1]{%
\savestack{\tmpbox}{\stretchto{%
  \scaleto{%
        \scalerel*[\widthof{\ensuremath{#1}}]{\kern-.6pt\bigwedge\kern-.6pt}%
                  {\rule[-\textheight/2]{1ex}{\textheight}}%
                              }{\textheight}%
}{0.5ex}}%
\stackon[1pt]{#1}{\tmpbox}%
}

\usepackage[hang,flushmargin]{footmisc}

\renewcommand{\epsilon}{\ensuremath\varepsilon}

\renewcommand{\phi}{\ensuremath{\varphi}}

\usepackage[linesnumbered,ruled,vlined]{algorithm2e}
\SetAlFnt{\scriptsize}
\SetAlCapFnt{\scriptsize}
\SetAlCapNameFnt{\scriptsize}

\newif\iftr
\newif\ifcnf
\newif\ifconf
\newif\ifbd

\newif\ifnohl

\newif\ifkdd
\newif\ifkddr
\newif\iffull
\newif\ifdebug
\newif\ifreg
\newif\iffullapp
\newif\ifacm
\newif\ifcr

\trfalse
\bdfalse
\crtrue
\cnftrue
\conftrue

\nohltrue

\conftrue
\regtrue
\debugtrue
\fulltrue
\kddtrue
\kddrtrue
\fullappfalse

\acmfalse

\sqfalse
\sqCAPfalse
\sqENfalse
\sqVSfalse

\ifcr
\fi

\RequirePackage[
  datamodel=acmdatamodel,
  style=acmnumeric,
  sorting=none,
]{biblatex}

\begin{document}

\date{}

\title{A High-Performance Design, Implementation, Deployment,\\ and Evaluation of The Slim Fly Network\vspace{-0.65cm}} %

\newcommand*\samethanks[1][\value{footnote}]{\footnotemark[#1]}

\ifbd
\else
\ifcr
\author{\rm{Nils Blach}\textsuperscript{\rm 1}, 
\rm{Maciej Besta}\textsuperscript{\rm 1},
\rm{Daniele De Sensi}\textsuperscript{\rm{1,2}},
\rm{Jens Domke}\textsuperscript{\rm 3}, \\
\quad \quad \rm{Hussein Harake}\textsuperscript{\rm 5},
\rm{Shigang Li}\textsuperscript{\rm{1,4}},
\rm{Patrick Iff}\textsuperscript{\rm 1},
\rm{Marek Konieczny}\textsuperscript{\rm 6},
\rm{Kartik Lakhotia}\textsuperscript{\rm 7}, \quad \quad \\
\rm{Ales Kubicek}\textsuperscript{\rm 1},
\rm{Marcel Ferrari}\textsuperscript{\rm 1},
\rm{Fabrizio Petrini}\textsuperscript{\rm 7},
\rm{Torsten Hoefler}\textsuperscript{\rm 1}\\
\vspace{-0.45cm}
\and
\textsuperscript{\rm 1} ETH Z{\"u}rich \quad
\textsuperscript{\rm 2} Sapienza University of Rome \quad
\textsuperscript{\rm 3} RIKEN Center for Computational Science (R-CCS) \\
\textsuperscript{\rm 4} BUPT, Beijing \quad
\textsuperscript{\rm 5} Swiss National Supercomputing Centre (CSCS) \quad
\textsuperscript{\rm 6} AGH-UST \quad
\textsuperscript{\rm 7} Intel Labs
\vspace{0.07cm}
\and
{\rm\{ nils.blach, maciej.besta, htor \} @ inf.ethz.ch}
}
\newcommand{\shortauthors}{Blach et al.}
\else
\author{\rm{Nils Blach}\textsuperscript{\rm 1}, 
\rm{Maciej Besta}\textsuperscript{\rm 1},
\rm{Daniele De Sensi}\textsuperscript{\rm{1,2}},
\rm{Jens Domke}\textsuperscript{\rm 3}, \\
\quad \quad \rm{Hussein Harake}\textsuperscript{\rm 5},
\rm{Shigang Li}\textsuperscript{\rm{1,4}},
\rm{Patrick Iff}\textsuperscript{\rm 1},
\rm{Marek Konieczny}\textsuperscript{\rm 6},
\rm{Kartik Lakhotia}\textsuperscript{\rm 7}, \quad \quad \\
\rm{Ales Kubicek}\textsuperscript{\rm 1},
\rm{Marcel Ferrari}\textsuperscript{\rm 1},
\rm{Fabrizio Petrini}\textsuperscript{\rm 7},
\rm{Torsten Hoefler}\textsuperscript{\rm 1}\\
\vspace{-0.45cm}
\and
\textsuperscript{\rm 1} ETH Z{\"u}rich
\and
\textsuperscript{\rm 2} Sapienza University of Rome
\and
\textsuperscript{\rm 3} RIKEN Center for Computational Science (R-CCS)
\and
\textsuperscript{\rm 4} BUPT, Beijing
\and
\textsuperscript{\rm 5} Swiss National Supercomputing Centre (CSCS)
\and
\textsuperscript{\rm 6} AGH-UST
\and
\textsuperscript{\rm 7} Intel Labs
\vspace{0.08cm}
\and
{\rm\{ nils.blach, maciej.besta, htor \} @ inf.ethz.ch}
}
\newcommand{\shortauthors}{Blach et al.}
\fi
\fi

\twocolumn[{%
\renewcommand\twocolumn[1][]{#1}%
\vspace{-0.48cm}
\maketitle
\ifbd
\vspace{-3.5cm}
\begin{center}
\Large \emph{Submitted to the Operational Systems Track}
\end{center}
\vspace{0.1cm}
\fi
\ifcr
\vspace{-0.66
cm}
\fi
\begin{center}
  \centering
    \captionsetup{type=figure}
\includegraphics[width=1\textwidth]{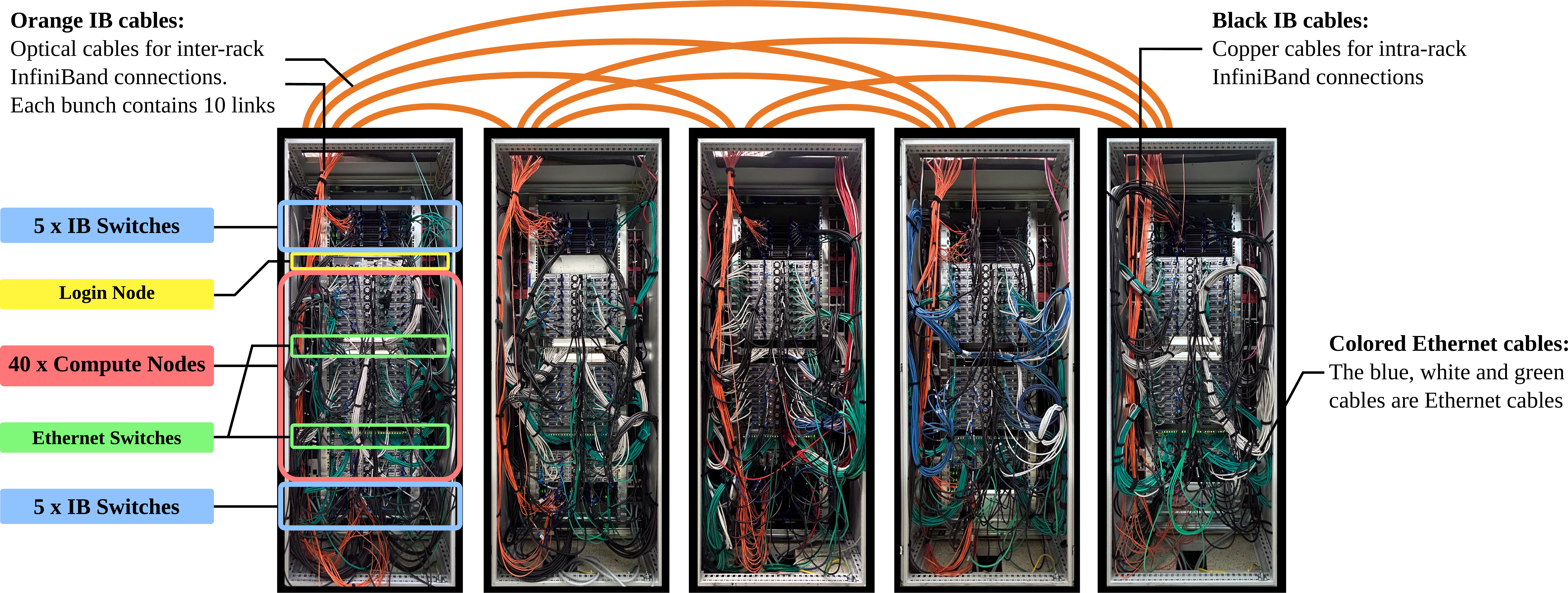}

\captionof{figure}{\textmd{First real-world deployment of the Slim Fly topology. The left-most rack displays labels detailing the arrangement of various components such as InfiniBand (IB) switches, compute nodes and Ethernet switches. Two types of IB links are present: black copper links for intra-rack connections and orange optical fiber links for inter-rack connections. The orange lines above the racks represent bundles of ten optical fiber links each. Additionally, blue, white and green (arbitrary color scheme) Ethernet cables are visible within the racks, which establish the cluster management network together with the Ethernet switches.}}
\vspaceSQ{-1em}
\label{fig:cluster-all-racks}
\end{center}%
}]

\begin{abstract}
Novel low-diameter network topologies such as Slim Fly (SF) offer significant cost
and power advantages over the established Fat Tree, Clos, or Dragonfly. To
spearhead the adoption of low-diameter networks, we design, implement, deploy, and
evaluate the first real-world SF installation.
We focus on deployment, management, and operational aspects of our test cluster with 200 servers and carefully analyze performance. 
We demonstrate techniques for simple cabling and cabling validation as well as a novel high-performance routing architecture for InfiniBand-based low-diameter topologies. 
Our real-world benchmarks show SF's strong performance for many modern workloads such as deep neural network training, graph analytics, or linear algebra kernels.
SF outperforms non-blocking Fat Trees in scalability while offering comparable or better performance and lower cost for large network sizes.
Our work can facilitate deploying SF while the associated (open-source)\footnote{\url{https://github.com/spcl/opensm}} routing
architecture is fully portable and applicable to accelerate any low-diameter
interconnect. 
\end{abstract}

\section{INTRODUCTION}

\emph{Low-diameter}\footnote{Network diameter is the maximum distance between
any two switches.} network topologies such as Slim
Fly (SF)~\cite{besta2014slim} have gained significant traction during the last
decade. Initial designs in that line of work, Dragonfly (DF)~\cite{dally08} and
Flattened Butterfly~\cite{dally07}, both with diameter three, focused on
improving latency and physical layout. After that, SF lowered the
diameter to two, based on an observation that low-diameter does not
only improve performance by reducing end-to-end latencies, but \emph{it also reduces
cost and power consumption}. This is because, when the diameter is lower,
packets on average traverse fewer switches, switch buffers, and links. Thus,
fewer links and buffers are needed to construct the network (for a
fixed bandwidth), and less dynamic power is needed for moving the packets
through the network.

SF's construction costs, consumed power, and latency are lower than those of Clos and Fat Tree (FT) by respectively, $\approx$25-30\%, $\approx$25-30\%, and $\approx$50\%~\cite{besta2014slim}. However, SF has still not seen a real physical deployment, and it is uncertain how to deploy SF in practice.
To spearhead the practical development of low-diameter networks and show the
state-of-the-practice, we design,
implement, deploy, and evaluate the first SF installation that includes switches and endpoints, 
as shown in Fig.~\ref{fig:cluster-all-racks}.
We discuss the encountered challenges, and 
we show that the construction process is straightforward and comparable
to established designs such as Clos. 

Moreover, to maximize performance benefits from using SF, we design and
implement a novel high-performance multipath routing scheme for general
low-diameter networks, and we install and use it with the deployed SF cluster.
Our routing shows superior performance over the state-of-the-art,
and it is independent of the underlying topology details and of the
interconnect architecture. Thus, it could be portably used on different
topologies (e.g., Xpander~\cite{valadarsky2015}) and on different architectures
(e.g., Ethernet or InfiniBand~\cite{IBAspec}).

\begin{figure*}[t]
\vspaceSQ{-1em}
\vspace{-0.1em}
\centering
\includegraphics[width=0.87\textwidth]{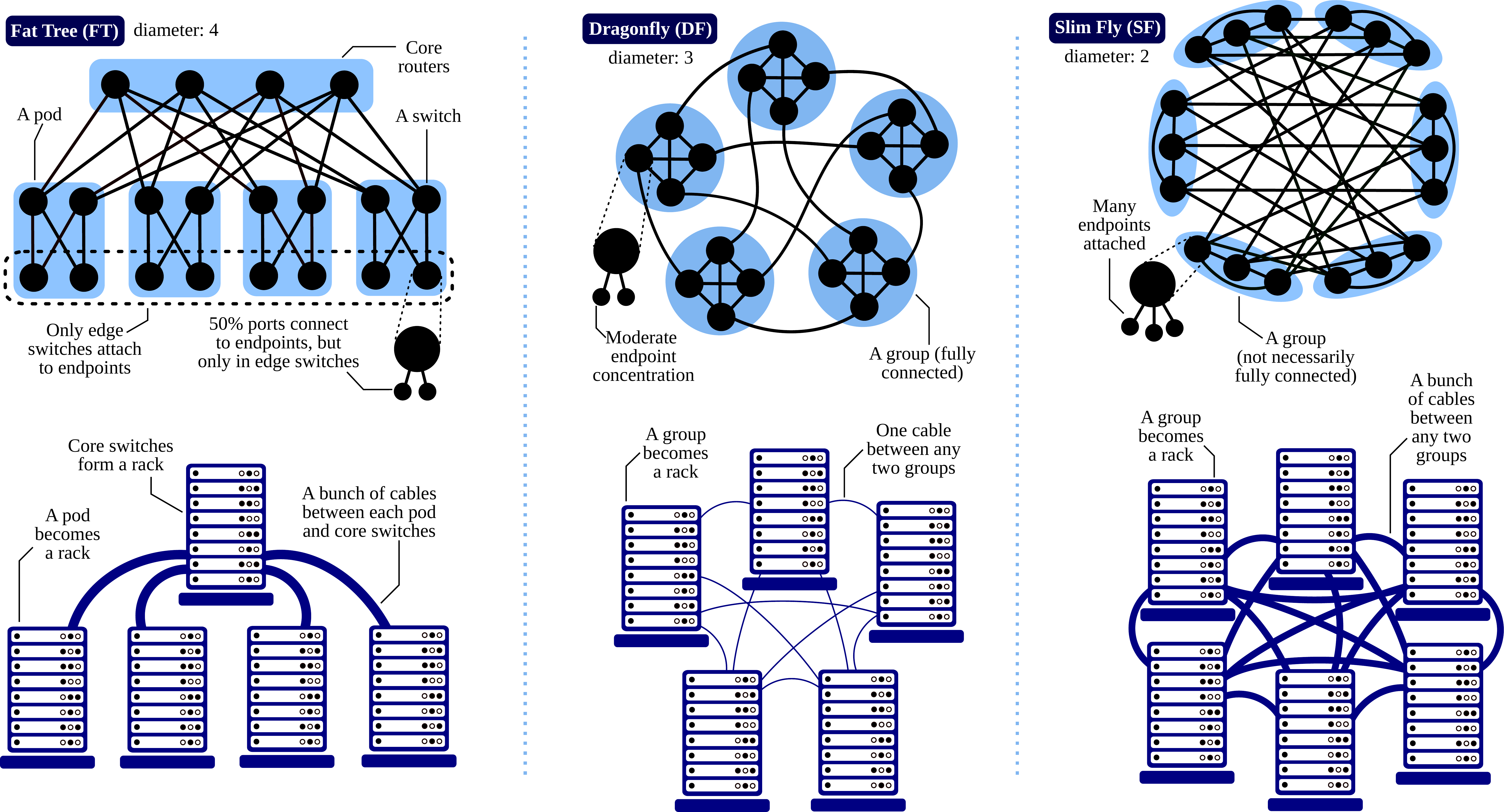}
\vspaceSQ{-1.5em}
\vspace{-0.6em}
\caption{\textmd{The structure of a small example Fat Tree (FT), Dragonfly (DF), and Slim Fly (SF), and the
corresponding installations. Each topology comes with a modular design, where switches form
groups (SF, DF) or pods (FT). Such groups can become racks in a physical installation.}}
\vspace{-1em}
\vspaceSQ{-1em}
\label{fig:nets}
\end{figure*}

The equipment available to us is based on the
InfiniBand (IB) architecture~\cite{IBAspec}. IB enables a high-speed switched
fabric with hardware (HW) support for remote direct memory access
(RDMA)~\cite{gerstenberger2013enabling, di2019network}. 
IB is widely used in
high-performance systems, for example four out of ten most powerful systems in
the Top500 list (Jun.~2023 issue)~\cite{dongarra1997top500}, manufactured by
IBM, Nvidia, and Atos, use the IB interconnect.
We use our routing protocol with the IB networking stack; our whole
implementation is publicly available to foster future research into
multipath routing. Importantly, we provide the first multipathing
for IB that can use arbitrary paths (including non-minimal and disjoint ones)
  and that is independent of the structure details of the underlying
  network~\cite{besta2020highr, domke_hyperx_2019}.

\sloppy
In our evaluation, we consider a broad range of communication-intense
applications that represent traditional dense computations (like physics simulations), 
sparse graph processing~\cite{besta2017push, besta2019demystifying, sisa, gms, besta2020high, besta2019practice}, 
deep neural network (DNN) training~\cite{ben2019modular, besta2022parallel, besta2023high}, 
and a number of microbenchmarks testing particular popular communication patterns. 
Our results showcase that SF delivers high performance while achieving optimal, or near optimal scalability, which directly translates to low construction costs.
To further reinforce these outcomes,
we also conduct a comprehensive comparison between SF and a non-blocking FT that we deploy using the same hardware.
Here, SF offers comparable or better performance to FT in a majority of used applications. 
Simultaneously, its superior scalability ensures up to 50\% cost improvements over FT, particularly for large installation sizes~\cite{besta2014slim}.

\section{NETWORK MODEL \& TOPOLOGIES}

We start with fundamental concepts and notation.
We model a network as an undirected graph $G = (V,E)$; $V$ is a set of
switches\footnote{\scriptsize We abstract away HW details and
denote switches and routers with a common term ``switch''. However, we use a
term ``routing'' when referring to determining a path, because IB switches in
our physical implementation have routing capabilities.} 
\iftr
\else 
\ifcr
($|V| = N_r$) and
\else
($|V| = N_r$ (number of switches)) and
\fi
\fi
$E$ is a set of full-duplex inter-switch cables (we do not model endpoints explicitly). A
network has $N$ endpoints, with $p$~endpoints attached to each switch
(\emph{concentration}). We also use the term \emph{node} to refer to either
a switch or any of its endpoints, when the discussion is generic. Total port count in a switch (\emph{radix}) is $k = k'
+ p$, where $k'$ is the number of channels from a switch to other switches
(\emph{network radix}). The diameter is $D$.
\ifcr
All the symbols are listed in Tab.~\ref{tab:symbols}.
\fi

\ifcr
\begin{table}[h]
\vspaceSQ{-1.5em}
\vspace{-0.5em}
\caption{The \textbf{most important symbols} used in this work.}
\centering
\scriptsize
\setlength{\tabcolsep}{2pt}
\ifsq\renewcommand{\arraystretch}{0.7}\fi
\vspace{-1.5em}
\begin{tabular}{@{}ll@{}}
\toprule
                   $V, E$ & Sets of vertices/edges (switches/links, $V=\{0,\dots,N_r-1\}$).\\
                   $N$& The number of endpoints in the network.\\
                   $N_r$& The number of switches in the network ($N_r = |V|$).\\
                   $p$& The number of endpoints attached to a switch.\\
                   $k'$& The number of channels from a switch to other switches.\\
                   $k$&\emph{Switch radix} ($k = k' + p$).\\
                   $D, d$&Network diameter and the average path length.\\
\bottomrule
\end{tabular}
\vspace{-1.0em}
\label{tab:symbols}
\end{table}
\fi

We overview SF's structure in Fig.~\ref{fig:nets}, and 
compare it to a 3--level Fat Tree
with diameter four, as they are widely used in medium and large
installations~\cite{al2008scalable, niranjan2009portland}, and to a diameter-3 Dragonfly,
which has also been deployed in practice~\cite{aries, slingshot-desensi}.
SF has $>$50\% fewer switches and $>$55\% fewer cables than a full-bandwidth
non-blocking FT of a comparable size.
Second, SF's switches form \emph{groups} that are not necessarily fully
connected; FT's edge and aggregation switches form \emph{pods}, DF's groups
are fully connected.
Third, both SF and DF are \emph{direct} topologies (each switch is attached to some
number of servers), while in a FT, only edge switches attach to servers.

\section{FIRST AT-SCALE SF INSTALLATION}
\label{sec:sf-real}
\begin{figure}[t]
\vspace{-0.2em}
\vspaceSQ{-1em}
\centering
\includegraphics[width=1.0\linewidth]{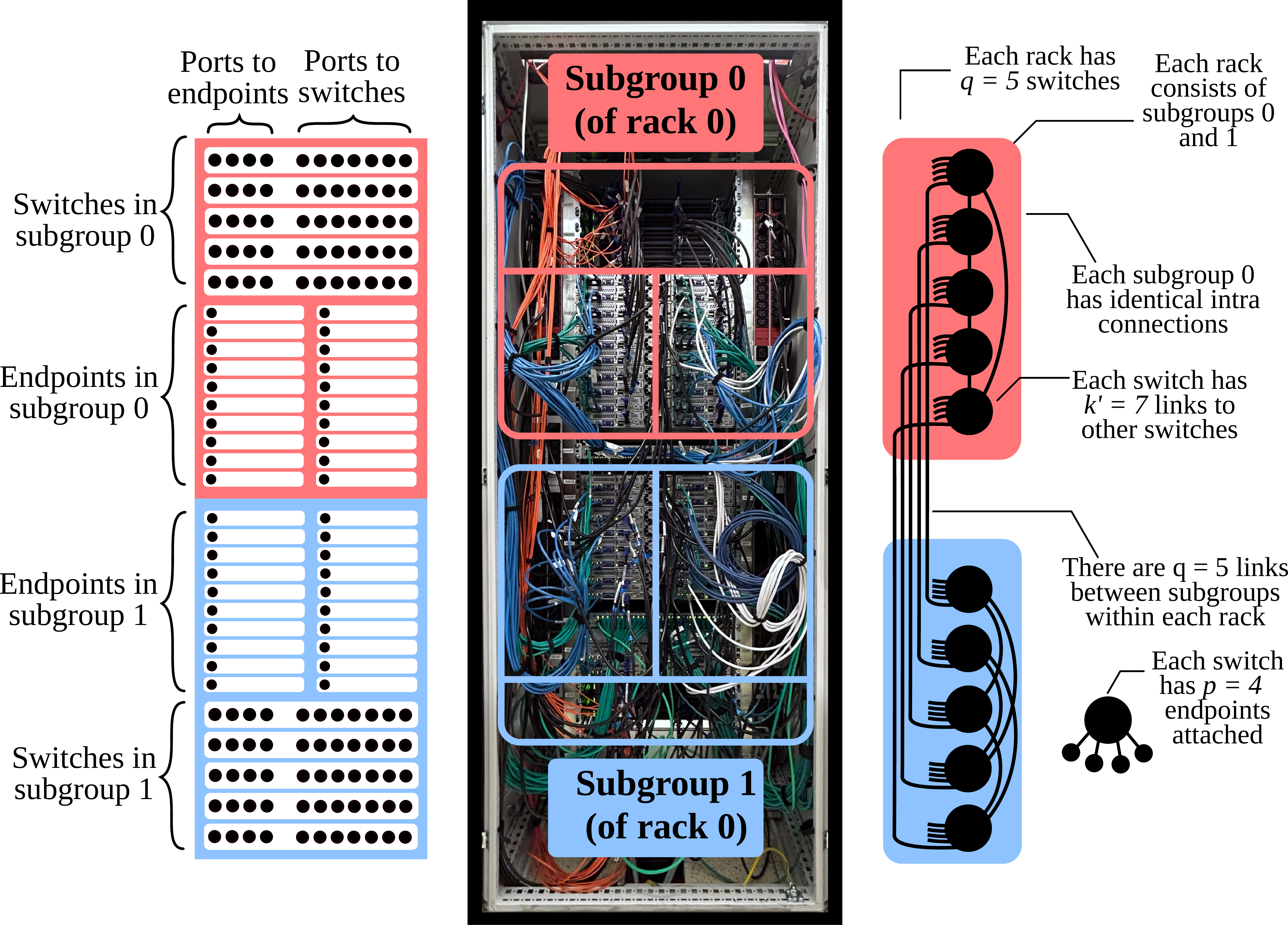}
\vspaceSQ{-1.5em}
\vspace{-1.5em}
\caption{\textmd{Internal organization of a rack. The image displays a side-by-side comparison of a theoretical diagram and an actual photograph of a single rack in the cluster. The rack consists of two distinct subgroups, each housing 5 IB switches and 40 compute nodes (endpoints). Each IB switch is connected to 4 endpoints and 7 other IB switches.}}
\vspace{-2em}
\vspaceSQ{-1em}
\label{fig:single-rack}
\end{figure}

We start by discussing the deployment of the first SF cluster, illustrating the simplicity
of its construction and arguing why deploying other SFs would also be
straightforward.
\ifbd
The cluster is hosted by [omitted for blind review].
\else
The cluster is hosted by the Swiss National Supercomputing Centre
(CSCS).
\fi

\subsection{Deployed Hardware Equipment}

We use $50$ $36$-port, $56$Gb/s IB SX6036 switches
and $200$ compute endpoints. Each endpoint hosts two $20$-core Intel Xeon CPUs
and $32$ GiB RAM, split equally in a Non-Uniform Memory Access (NUMA)
configuration, and a single Mellanox ConnectX-3 MT4099 HCA, which implements
the IB Architecture Specification Volume 1, Release 1.2. Copper and optical
cables are used for intra and inter-rack switch connections, respectively.

\subsection{Topology Structure and Construction}

We use a SF based on the graphs by McKay, Miller, and Širáň~\cite{mckay1998note}. 
We outline its structure, the details are
in~\cref{sec:sf_cons} and in the original SF paper~\cite{besta2014slim}.
The complete SF installation is shown in Fig.~\ref{fig:cluster-all-racks}
with a highlighted view of the group structure in Fig.~\ref{fig:single-rack}.
One first chooses a prime power $q$; $q$ is an input parameter that
determines the whole topology structure. For example, the number of vertices
(switches) is $N_r = 2 q^2$ and the network radix $k' = \frac{3q-\delta}{2}$. 
In our case, $N_r = 50$, thus $q = 5$ and $k' = 7$ (every switch
connects to $7$ other switches).
Interestingly, this construction forms the famous Hoffman-Singleton
graph~\cite{hoffman1960moore, hafner2003hoffman}, which is \emph{optimal} with
respect to the Moore Bound~\cite{MooreBound}.
Finally, one uses $p = \ceil*{\frac{k'}{2}}$ endpoints connected to each switch to ensure
\emph{full global bandwidth}~\cite{besta2014slim}. In our case, $p = 4$.
Note that, while the switch port count in the considered SF is $k'+p = 11$
(and 11-port switches would be the appropriate selection when building the
SF from scratch),
we use 36-port switches because this has been the only HW equipment
available to us.

The whole installation consists of five identical racks. Every two racks are
connected with the same number of $2q = 10$ cables. There are $2q = 10$
switches in each rack.
Each rack consists of two \emph{subgroups}, \emph{subgroup~0} and
\emph{subgroup~1}. All subgroups~0 and all subgroups~1 are identical, but a subgroup~0 and~1
are usually different. We place switches
from subgroup~0, together with their attached endpoints, at the top of each
rack; subgroup~1 goes to the bottom of the rack.
The details on how any two switches are connected is determined by the
underlying algebraic structure of the SF topology. 
We offer full details in~\cref{sec:sf_cons},
with~\cref{sec:sf_cons_eq} explaining the three simple equations that determine switch connectivity; 
here, we stress that the deployment is straightforward.

\subsection{Deployment Efficiency and Ease}

\begin{figure}[h]
\vspaceSQ{-1em}
\vspace{-0.6em}
\centering
\includegraphics[width=0.9\columnwidth]{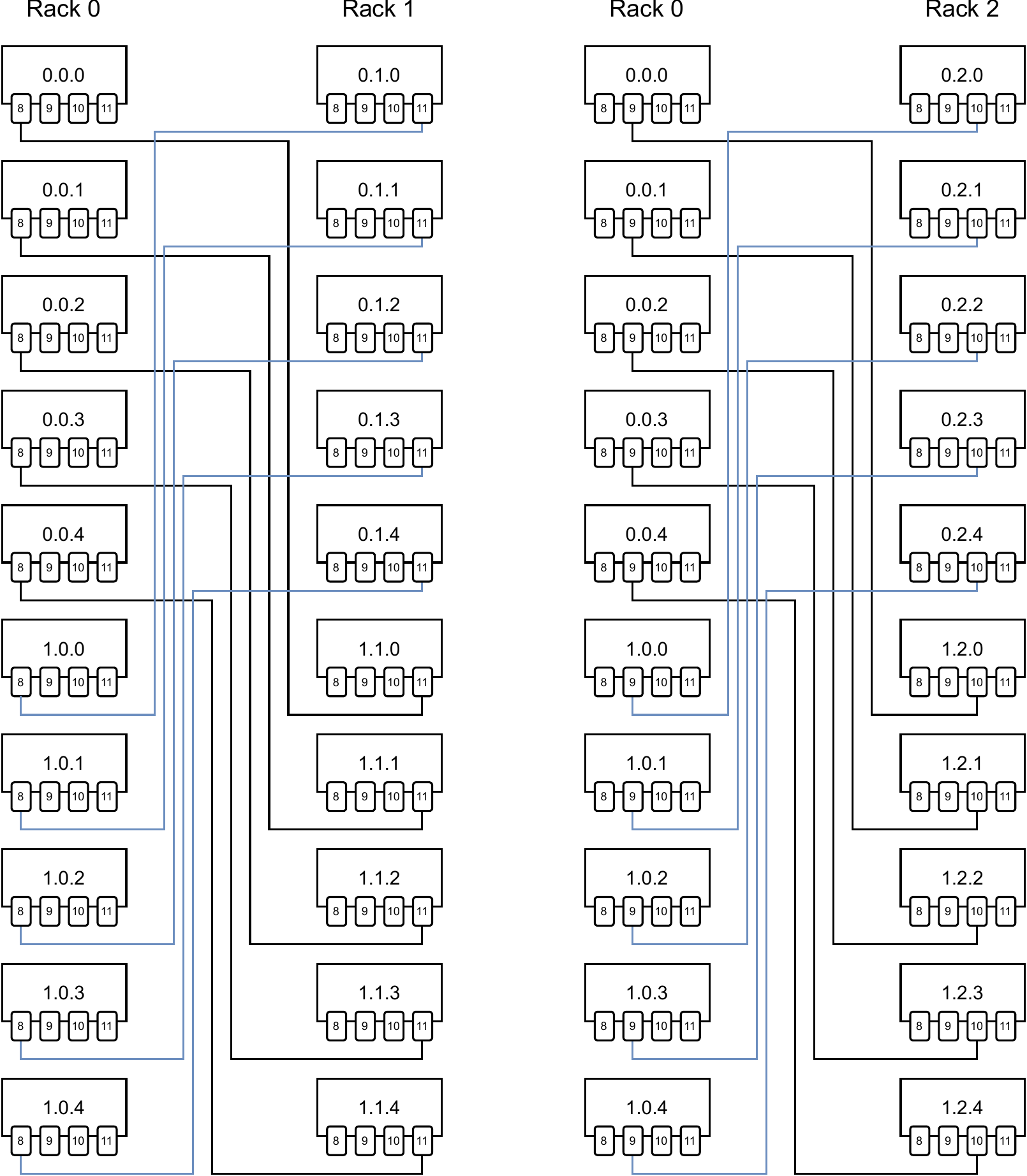}
\vspaceSQ{-1.5em}
\vspace{-0.6em}
\caption{\textmd{Illustration of the example diagrams created from the output of our scripts,
facilitating the cabling process. The diagrams show all the inter-rack
connections and the corresponding ports in switches. Each switch is labeled
using a triple $(S,R,I)$, where $S \in \{0,1\}$ indicates the subgroup type, $R
\in \{0, ..., 4\}$ indicates the rack, and $I \in \{0, ..., 4\}$ is the
consecutive switch ID within a rack/subgroup. Then, we only show ports 8--11;
these ports are used to connect racks. Ports 1--4 (for endpoints) and 5--7 (for
intra-rack switch-switch links) are omitted for clarity. 
The equations presented in~\cref{sec:sf_cons_eq} determine which switches 
are connected based on the assigned labels.}}
\vspace{-1.0em}
\vspaceSQ{-1em}
\label{fig:diagrams}
\end{figure}

To facilitate deployment, we develop scripts that outline both
intra- and inter-rack connections. The output of these scripts 
can be used to create diagrams for every rack pair to ensure a smooth wiring process. 
Thanks to the algebraic structure of the SF
topology, such descriptions for any SF can be automatically generated, providing
concrete port-to-port link descriptions and rack placements for each switch.
We illustrate an example diagram of connections between racks 0 and 1, and between
0 and 2, that was created based on these generated descriptions, in Fig.~\ref{fig:diagrams}.

We use our scripts as a basis of an efficient 3-step wiring process.
First, we wire intra-subgroup connections; they are identical across all racks
for each of the two subgroups.
The second step consists of connecting each switch from subgroup~0 to its
neighboring switches in subgroup~1 within the same rack. As the subgroups are
of equal size, an incorrectly connected pair will result in easily recognizable
errors, which break that symmetry. Lastly, the inter-rack connections are
established. Hereby, the fact that each switch in a rack uses the same port to
connect to the switches in another rack, enables straightforward connection of
rack-pairs.

The simplicity of the wiring process can mainly be attributed to the scalable three-step 
approach, which is equally applicable to larger SF topologies, 
enabling the efficient deployment of SF clusters.
Overall, stripping the previous system and executing the 3-step wiring process 
were completed within 3 days by a team of two.

\subsection{Correctness Verification}

We provide a set of scripts that ensure the correctness of the cabling.
These scripts utilize the auto-generated port-to-port link descriptions
and rack placements for each switch and compare it with the output of 
\texttt{ibnetdiscover}, an IB command that performs fabric 
discovery. This allows us to not only identify incorrectly wired cables
and provide concrete instructions on how to rectify mistakes, but also
detect missing or broken links. These scripts could even be used
on a live cluster, while going through the wiring process, to immediately identify and 
flag errors.

\section{HIGH-PERFORMANCE MULTIPATHING}
\label{sec:routing-des}

We now propose a novel high-performance multipath routing protocol for
low-diameter networks, which we use on the described SF deployment.
For this, we extend the recently proposed FatPaths multipath routing protocol~\cite{besta2020fatpaths} so that it offers
vastly superior throughput while still ensuring very low latency.

\begin{figure}[t]
\vspace{0.5em}
\vspaceSQ{-1em}
\centering
\includegraphics[width=0.48\textwidth]{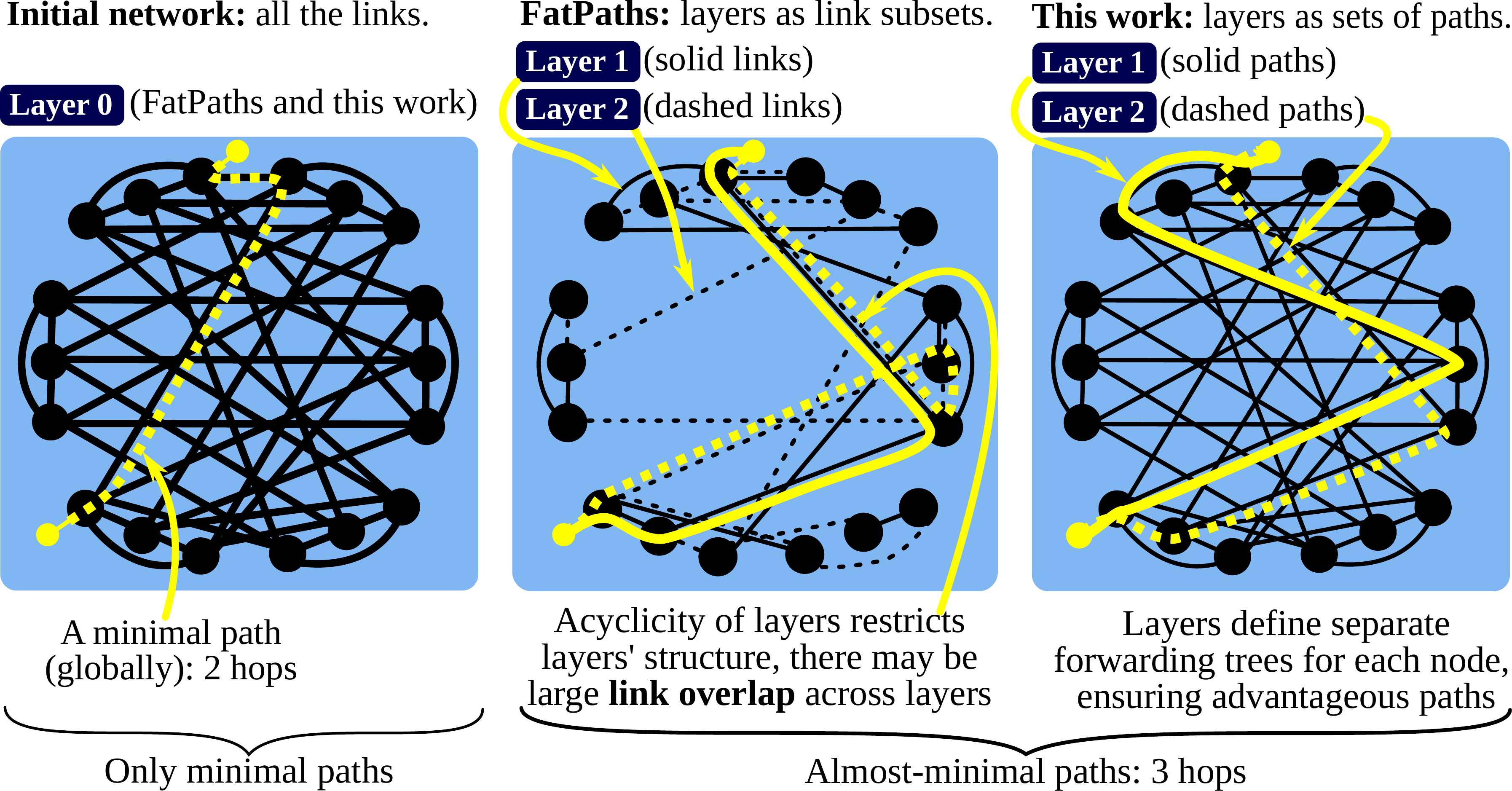}
\vspaceSQ{-2em}
\vspace{-1.5em}
\caption{\textmd{Layered routing in FatPaths and in this work.
Traffic is divided and sent using different layers.
Our scheme relaxes the requirement in FatPaths for all layers to be trees, 
as in our scheme deadlock resolution is decoupled from layer creation.
\textbf{This ensures more flexibility in developing layers, leading to more
throughput.} Specifically, while in FatPaths, paths in different layers often
overlap (cf.~Layer~1 and~2), 
our routing alleviates this issue and reduces overlap/congestion and increases performance.}}
\vspace{-1.2em}
\vspaceSQ{-1em}
\label{fig:layered}
\end{figure}

\subsection{Original FatPaths Routing in Slim Fly}

In terms of path diversity, FT has multiple same-length minimal paths between
any two edge switches. Thus, one often uses ECMP~\cite{hopps2000analysis} for
multipath routing in FT. In SF (and to some degree in 
DF~\cite{besta2020fatpaths}), there is usually only one minimal path, but
multiple ``almost'' minimal paths between any switch pair.
This makes it challenging to achieve high path diversity in SF using ECMP. 
To alleviate this and to enable
non-minimal high-speed multipathing in SF, the \emph{FatPaths architecture} has recently
been proposed~\cite{besta2020fatpaths}.
FatPaths harnesses the concept of \emph{layered
routing}~\cite{mudigonda2010spain, stephens2012past} for low-diameter networks.
In layered routing, one first creates \emph{layers}: subsets of switch-switch
links. Within one layer, one uses shortest-path routing. However, as a layer
does not contain all the links, paths within this layer are usually non-minimal
(in the global sense).
If two nodes\footnote{\scriptsize Multipathing can be applied both at the
switch and at the endpoint level. Thus, we use a term ``node'' to
  refer to switches or endpoints when a discussion is generic} want to
  communicate using multiple paths, the sending node simply sends its data
  using paths residing in different layers. 
Note that multipathing is orthogonal to transport-level issues, and one can
use different layers to transfer different flows between two nodes, but also
different packets or flowlets within one flow~\cite{besta2020fatpaths}.
In FatPaths, selecting links (when constructing layers) is done with simple
random uniform sampling; a more elaborate scheme minimizing load
imbalance is also provided.
Layered routing is summarized in Fig.~\ref{fig:layered}.

\subsection{Proposed Multipath Routing: Summary}

The central issue in layered routing is \emph{how to divide links into layers}.
We aim to minimize the number of layers (which minimizes the usage of HW
resources in switches) while simultaneously maximizing the number of disjoint
and almost-minimal paths between any switch pair (for more path diversity).
Moreover, a detailed analysis from FatPaths indicates that -- to maintain
high performance in layered routing in virtually all low-diameter networks 
and traffic patterns -- at least \emph{three} disjoint paths per switch
pair are needed~\cite{besta2020fatpaths}. Thus, the
main goal of the layer construction algorithm is to find a minimum set of
layers that together provide each switch pair with at least three disjoint
paths while ensuring minimum overlap between specific layers. Ideally, these
three paths include the minimal one (that always exists) and two ``almost''
minimal ones (in the following, an ``almost'' minimal path means a path
that is longer by one hop than the minimal path
between two given switches).

An overview of our proposed layer routing is shown in Fig.~\ref{fig:layered}
(right). The key difference between our scheme and FatPaths is that we do \emph{not}
remove links from layers in order to ensure deadlock-freedom or to introduce
non-minimal paths. Instead, we decouple deadlock resolution from layer creation,
and explicitly construct paths satisfying the appropriate
constraints on their count, non-minimal length, and well-balancedness.
This facilitates creating layers that result in \emph{much} higher throughput.

\subsection{Generating Routing Layers}
\label{sec:r-overview}

Our layer construction 
scheme is detailed in
Algorithm~\ref{algo::pseudo_layer}. The input is the topology
of inter-switch connections $G=(V,E)$, and the desired number of layers $|L|$.
The output is a set of layers $L$, where each layer contains a collection of paths 
connecting different pairs of nodes. These paths together define a separate 
forwarding tree for each node.

The layer generation starts with assigning all links to layer~1. In layer $1$,
we only use minimal paths, as we want to ensure that the single minimal path
existing between all node pairs is included in at least one layer for each
pair.
Moreover, a matrix~$W$ and a priority queue~$p$ are initialized. These
structures are used to find advantageous non-minimal paths for each node pair.
Intuitively, a priority~$p(u,v)$ of a node pair~$u,v$ is determined by the number of
non-minimal paths {already assigned to $u,v$} (and maintained in other layers).
The higher $p(u,v)$ is, the lower the priority of $u,v$ is. Hence, when looking
for new non-minimal paths, node pairs with fewer paths assigned are
  prioritized.  This facilitates balancing the number of advantageous paths
  across {all} pairs of nodes, to eliminate potential hotspots in the network.
 
Second, each entry~$W(r,s)$ in matrix~$W$ describes the weight of a link
between switches~$r,s$. This weight equals the number of paths (from any layer) that
already use this link. The higher~$W(r,s)$ is, the more paths use the
corresponding link. Hence, when selecting new paths, we use $W$ to balance
numbers of paths across single links, minimizing risk of congestion. 
We also use $W$ to balance the paths in the first layer to ensure minimal overlap of minimal paths.

Then, for every layer $2\dots|L|$, and for each node pair in each layer, we
find a single almost-minimal path that {minimizes overlap
with respect to paths already added to any other layer}. 
For this, when finding paths in a layer~$l$, we first copy the current
priorities of node pairs into a list that preserves the
current state of priorities (\emph{copy\_pairs}). 
Here, node pairs with the same priority are in a random order, but come before
any node pair with lower priority. Note that each node pair appears twice in
the list, once for each direction. This enables using different paths when
routing in different directions, further increasing the flexibility of path
selection.

After that, we iterate over each node pair, in an attempt to construct a path
for each such pair in each layer. Note that, in principle, it is possible that
one cannot find a path for each node pair in each layer (we elaborate on
dealing with such rare cases in~\cref{sec:layer_gen}; we resolve them with a simple 
fallback to a minimal path -- our evaluation
shows that this does not negatively impact throughput).

In each such iteration, we first use the \emph{find\_path} routine to try to
find an almost-minimal path for a given node pair $pair$, based on already
inserted paths for that layer (specified in $l$) and weights assigned
to each link (specified in $W$).
If we are able to find a valid path, we accordingly update priorities~$p$ (\emph{update\_priorities}) and
link weights~$W$ (\emph{update\_weights}).
Finally, we insert the path into layer~$l$ (\emph{add\_path\_to\_layer}).

\iftr
\else
\vspace{-1em}
\fi
\begin{algorithm}[]
\DontPrintSemicolon
\SetAlgoLined
\SetKwInOut{Input}{Input}\SetKwInOut{Output}{Result}
\Input{\ Network topology $G = (V,E)$, number of layers $|L|$}
\Output{\ A set of $L$ routing layers}
 \tcp*[l]{$W \in \mathbb{R}^{N_r \times N_r}$ contains weights of links; $p$ is a priority queue, with entries being pairs of nodes}
 $W$ = \textbf{init\_link\_weight\_matrix}() \tcp*[l]{Set all matrix entries to 0}
 $p$ = \textbf{init\_p\_queue}($G$) \tcp*[l]{Each node pair gets the same priority}
 $L$ = $\{E\}$ \tcp*[l]{Layer~0 contains all the links ($E$)}
 \For{$l = 1$ \KwTo $|L|-1$}{ 
 \textbf{init\_layer}($l$) \tcp*[l]{Initialize the next layer as empty}
 $node\_pairs$ = \textbf{copy\_pairs}($p$)\; 
 \While{$node\_pairs \neq \emptyset$}{
    $pair$ = $node\_pairs$.dequeue()\;
    $path$ = \textbf{find\_path}($G$, $W$, $pair$, $l$)\;
    \If{\textbf{valid}($path$)} {
        \textbf{update\_priorities}($path$, $p$)\;
        \textbf{update\_weights}($path$, $W$)\;
        \textbf{add\_path\_to\_layer}($path$, $G$, $l$)\;
    }
 }
 $L = L \cup \{l\}$ \tcp*[l]{Add a new layer to finalized layers}
}
\caption{Construct routing layers; details are in~\cref{sec:r-overview}}
\label{algo::pseudo_layer}
 
\end{algorithm}
\iftr
\else
\vspace{-1.5em}
\fi

\section{IMPLEMENTATION OF MULTIPATHING}
\label{sec:routing-impl}

The IB architecture~\cite{IBAspec} enables a high-speed switched
fabric with HW support for RDMA~\cite{gerstenberger2013enabling,
infiniband2014rocev2} and atomic operations~\cite{schweizer2015evaluating}. IB
provides lossless destination-based packet forwarding that relies on
link-level, credit-based flow control~\cite{Dally:2003:PPI:995703}.
We now discuss the used IB features.

An IB network usually forms a single subnet consisting of physical IB switches
and \emph{Host Channel Adapters (HCAs)} that correspond to Ethernet NICs. All 
communication up to and including the transport layer is implemented within 
these two components.

Routing configuration is managed by a centralized \emph{subnet manager
(SM)}. The SM configures connected IB devices, appropriately computes the
forwarding tables to implement the used destination-based routing algorithm, 
and monitors the network for failures.
Within an IB subnet, each HCA and each switch receive a unique \emph{local
identifier (LID)}, assigned by the SM.

Each physical IB port has several independent \emph{virtual
lanes (VLs)}. Each VL has its own receive and transmit buffers and flow control
resources. There can be up to 15 VLs per physical port (depending on the 
equipment) and 1 VL for management traffic.
Multiple VLs per port are used for deadlock freedom and to eliminate head-of-line
blocking~\cite{Dally:2003:PPI:995703} (we discuss deadlocks in more detail in~\cref{sec:df}).

Each switch provides a forwarding table called the \emph{Linear Forwarding
Table} (LFT) that -- for a given packet -- determines the outgoing port using
the destination address (DLID) from the packet header.
Then, for a given outgoing port, to determine the outgoing VL for a given
packet, the switch uses a four-bit Service Level (SL) field from the packet
header, in combination with the incoming and outgoing packet ports, to index into the
\emph{SL-to-VL table}.
This enables packets to change virtual lanes at each hop and it allows for
seamless utilization of switches with potentially different numbers of virtual
lanes.

\subsection{Routing}

OpenSM, our choice of IB compliant SM, provides complete subnet information, including a list containing all
nodes (switches, HCAs, routers) and ports, as well as the connections between
them. 
We use this information to create and populate forwarding tables so that they
implement the prescribed layered routing.

\textbf{Multipathing}
In ECMP, each router stores multiple possible next-hops that each lie on a
minimal path towards the destination. This approach of storing multiple
next-hops for a given destination is not possible in IB.  However, it can be
emulated by assigning multiple LIDs to each HCA,
a feature that we use to enable
multipathing and to implement our layered routing in an IB setting. An HCA can
receive a contiguous range of LID addresses. This range is determined by the so
called \emph{LID Mask Control} (LMC) value. Specifically, for an LMC equal $x$,
each HCA port hosts a consecutive range of $2^x$ LIDs.
Then, one routes
towards each such LID using a \emph{different} path. 
\iftr
\else
\ifcr
\else
However, due to the fact that LIDs have a fixed size of 16 bits, the address space is limited and one can not assign infinitely many LIDs to each HCA.
We perform an extensive analysis on the effect of the LMC value on the maximum network size in the appendix (Sec.~\ref{sec:max_net}).
\fi
\fi
We use the information provided by OpenSM to appropriately populate
forwarding tables so that they implement the layered routing described 
in~\cref{sec:routing-des}.

\textbf{Implementation of Layers}
We assign multiple addresses to each node; one address falls into one layer
(each layer gets one address from each node). 
Hence, a layer is physically formed by the assigned addresses and the
associated forwarding entries that route traffic to these addresses. The
forwarding entries are set according to the specification of layers in the
initialization phase.
Our scheme for constructing layers provides a data structure $port$, which
specifies the output port to be used for a packet traveling to a node $d$,
from a switch $s$, within a layer $l$; this output port is denoted with
$port[l][s][d]$.

\textbf{Routing Within Layers}
The number of layers equals the number of addresses assigned to each node.
Thus, we can treat the layer ID as the offset to the base (i.e., to the first)
LID of each node. Hence, for instance, routing in the first layer (ID $0$)
uses the base LID of each node, whereas routing in the second layer uses the
base LID plus offset~1.

\textbf{Populating Forwarding Tables}
To populate forwarding entries, we add a value $port[l][s][d]$ into the LFT of
switch~$s$, as the outgoing port number for packets being routed towards node~$d$.  As the
destination address, we use the base LID of the node, \emph{increased by the
offset}~$l$, to ensure routing within layer $l$. 
As the last step, we run a deadlock-resolution scheme that fills
all SL-to-VL tables, eliminating the risk of deadlocks (cf.~\cref{sec:df}).

\subsection{Deadlock-Freedom}
\label{sec:df}

One downside of IB's credit-based flow control ensuring losslessness is the
possibility of \emph{deadlocks}. Specifically, an IB network may enter a state in
which packets in different buffers wait for each other indefinitely long to
free the buffers, resulting in a deadlock. To overcome this, most routing
schemes use different VLs to send packets~\cite{nue_routing, domke-deadlock-2011, 
schneider2016ensuring, Shim_2009, skeie2004lash, skeie2002layered}. By splitting a
single port buffer into multiple independent logical VLs, one can break
dependencies between waiting packets.

In FatPaths, each layer is {acyclic}, to ensure no deadlocks within
each layer. 
However, this does not imply global deadlock-freedom on IB
because of its lossless design based on channels. Specifically, one has to
ensure that dependencies between packets using routes stored in \emph{any}
layers are also deadlock-free.
Thus, we change the FatPaths approach by decoupling deadlock-avoidance
from layer creation. Instead, we apply deadlock-removal \emph{after} the layers are created.
This also enables much more throughput because acyclic layers vastly
restrict the choice of paths to be taken.

In our IB implementation, we propose and enable the use of two different deadlock-avoidance schemes. 
Firstly, if a sufficient number of VLs is available, we use the 
scheme introduced with the Deadlock-Free Single Source Shortest-Path 
(DFSSSP)~\cite{domke-deadlock-2011} algorithm, which is already integrated in IB.
Intuitively, given a ready routing (i.e., the populated forwarding tables),
DFSSSP first finds all dependencies that could lead to a deadlock, and then it
iteratively accommodates these dependencies in a deadlock-free way, by
assigning selected routes to use yet unoccupied VLs.
If not enough VLs are available, the algorithm fails. If not all VLs are
exhausted, DFSSSP additionally balances the number of paths using each VL, for
more throughput.

By increasing the number of layers used, the total number of unique paths between 
node pairs increases, resulting in a higher number of virtual lanes (VLs) required 
to resolve deadlocks using the DFSSSP scheme. 
To maximize the number of supported layers, we propose a novel deadlock avoidance scheme
based on the Duato's approach \cite{Duato:2002:INE:572400}, that 
is agnostic to the number of layers and tailored for IB deployments that rely exclusively 
on paths of length $<=3$, such as those based on SF with our multipath routing method.
The proposed algorithm ensures that the first, second, and third inter-switch hop of any path 
connecting two nodes use disjoint subsets of VLs. To achieve this, at least three VLs need 
to be available, and switches, for a given packet, must be able to identify their respective 
positions on the path using only the packet's SL, incoming and outgoing port.

To illustrate the algorithm's functionality, we consider each case individually. 
The first case, which involves paths of length 1 ($sw_1 - sw_2$), can be solved trivially 
since $sw_1$ can determine that it is the first hop along the path by checking whether the 
incoming packet port is connected to an endpoint. 
This information can then be encoded easily in the SL-to-VL table.

The strategy to address the second case, paths of length 2 ($sw_1 - sw_2 - sw_3$), 
is the same as the one for case three; therefore, we only present it once.
In the third and final case, paths of length 3 ($sw_1 - sw_2 - sw_3 - sw_4$), we treat $sw_1$ 
as in case one but use a different approach to differentiate between $sw_2$ and $sw_3$.
We establish a proper coloring of switches, using at most as many colors as there are available SLs.
This color assignment is then mapped to SLs, ensuring each switch has a unique color 
and SL among its neighbours.
By setting the SL of a packet routed along a path of length 2 or 3 to the SL assigned to the second
switch ($sw_2$) along that path, it is guaranteed that the packet's assigned SL matches the SL of the second hop
but not the SL of the third. 
Subsequently, if a switch is neither the first nor last hop on a path -- a condition trivially determined 
through the incoming and outgoing packet ports -- then the switch's position along the path can be ascertained by 
whether the incoming packet's assigned SL matches the SL assigned to the switch. 
Specifically, if the SLs match, then the given switch must be the second hop; if they don't, then it must be the third.
Thus, we can differentiate the second hop from a potential third hop and select 
the appropriate subset of VLs at each hop accordingly.

If fewer than $3$ VLs are available or no proper coloring using the available SLs can be established,
the algorithm fails. Similar to the DFSSSP scheme, the disjoint VL subsets can be chosen to balance the
number of paths crossing each VL.

\subsection{Load Balancing}

For load balancing, we rely on the respective protocol higher up in the stack
to choose a layer out of the set of possible ones available for a given
destination. In our case, this is the Open MPI~\cite{gabriel2004open}
implementation of the Message Passing Interface (MPI)
standard~\cite{clarke1994mpi}. Open MPI serves as a communication library and
directly interfaces with the IB networking API (Verbs). To optimize traffic flow, 
we utilize Open MPI's default load balancing technique, 
which distributes traffic evenly across the available paths using a round-robin selection process. 
More advanced, adaptive schemes can seamlessly be used by changing the selection policy.

For fault tolerance, we rely on IB's subnet manager.
We stress that our routing can be seamlessly used with other transport schemes
besides the ones used in the deployed cluster.

\ifcr
\subsection{Path Diversity vs.~Network Size}

Increasing the number of different paths between each node pair
requires more layers and thus also more
addresses assigned to each node (i.e., a larger LMC value).
However, using more addresses within one node decreases the maximum number of
nodes that can be used in the network overall (because the address field size
is fixed to 16 bits).
We analyze this tradeoff in Tab.~\ref{tab:uff}. We assume the maximum SF network based
on \{36, 48, 64\}-port switches, that guarantees full global bandwidth. The results
illustrate that one can use $4$ layers without having to make any compromises
on the networks size, but anything beyond $4$ layers would reduce the maximum
network size. At this point, the constraining factor is no longer the switch radix,
but the address space.
In~\cref{theory} and~\cref{sec:eval}, we show that -- fortunately -- our
routing scheme's performance is already quite substantial with just 4 layers and 
does not need more than 8 layers for high performance.

\begin{table}[h]
\vspace{-0.5em}
\caption{\textmd{Maximum number of switches and servers supported by a
single-subnet, full global bandwidth, SF-based IB network, with
$\textbf{``\#A''}=2^{LMC}$ many addresses per node.}} 
\centering
\setlength{\tabcolsep}{4pt}
\scriptsize
\begin{tabular}{lllllllllllll}
\toprule
    \multicolumn{1}{c}{} & \multicolumn{4}{c}{\textbf{36-port switches}} & \multicolumn{4}{c}{\textbf{48-port switches}} & \multicolumn{4}{c}{\textbf{64-port switches}} \\
    \textbf{\#A} & $N_r$ & $N$ & $k'$ & $p$ & $N_r$ & $N$ & $k'$ & $p$ & $N_r$ & $N$ & $k'$ & $p$\\ \midrule
    1 & 512 & 6144 & 24 & 12 & 882 & 14112 & 31 & 16 & 1568 & 32928 & 42 & 21\\
    2 & 512 & 6144 & 24 & 12 & 882 & 14112 & 31 & 16 & 1250 & 23750 & 37 & 19\\
    4 & 512 & 6144 & 24 & 12 & 800 & 12000 & 30 & 15 & 800 & 12000 & 30 & 15\\
    8 & 450 & 5400 & 23 & 12 & 450 & 5400 & 23 & 12 & 450 & 5400 & 23 & 12\\
    16 & 288 & 2592 & 18 & 9 & 288 & 2592 & 18 & 9 & 288 & 2592 & 18 & 9\\
    32 & 162 & 1134 & 13 & 7 & 162 & 1134 & 13 & 7 & 162 & 1134 & 13 & 7\\
    64 & 98 & 588 & 11 & 6 & 98 & 588 & 11 & 6 & 98 & 588 & 11 & 6\\
    128 & 72 & 360 & 9 & 5 & 72 & 360 & 9 & 5 & 72 & 360 & 9 & 5\\
\bottomrule
\end{tabular}%
\vspace{-1.0em}
\label{tab:uff}
\end{table}
\fi

\begin{figure*}[t]
\vspaceSQ{-1em}
\centering
\includegraphics[width=1.0\textwidth]{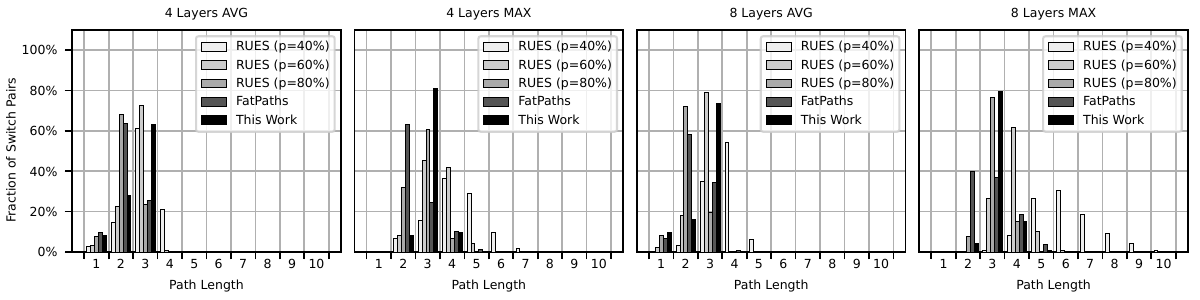}
\vspaceSQ{-2em}
\vspace{-2.5em}
\caption{\textmd{Histograms of average path lengths and maximum path lengths across all layers for each switch pair.}}
\vspace{-1em}
\vspaceSQ{-1em}
\label{fig:theory-pathlengths}
\end{figure*}

\begin{figure}[t]
\centering
\includegraphics[width=1.0\columnwidth]{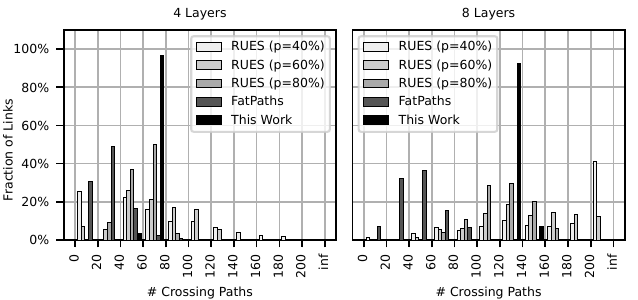}
\vspaceSQ{-2em}
\vspace{-2em}
\caption{\textmd{Histograms (bin size = 20) of counts of paths crossing each individual link.}}
\vspace{-1em}
\vspaceSQ{-1em}
\label{fig:theory-crossing}
\end{figure}

\begin{figure}[t]
\vspace{-0.5em}
\centering
\includegraphics[width=1.0\columnwidth]{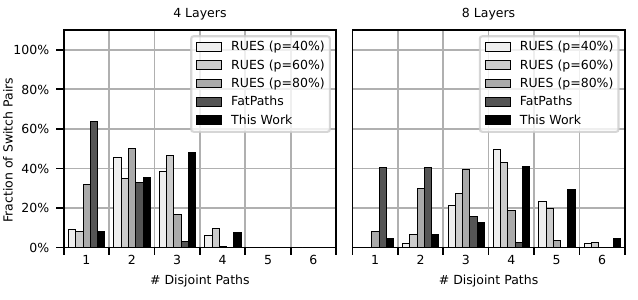}
\vspaceSQ{-2em}
\vspace{-2.5em}
\caption{\textmd{Histograms of counts of disjoint paths for different switch pairs.}}
\vspace{-1.5em}
\vspaceSQ{-1em}
\label{fig:theory-disjoint}
\end{figure}

\begin{figure}[t]
\centering
\includegraphics[width=1.0\columnwidth]{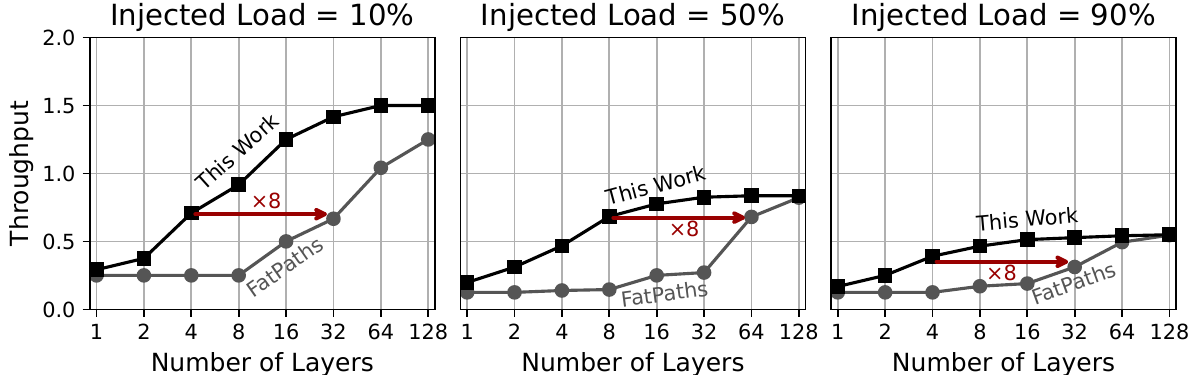}
\vspaceSQ{-2em}
\vspace{-1.7em}
\caption{\textmd{Maximum achievable throughput for the adversarial traffic pattern under three different injection loads (fraction of communicating endpoint pairs).}}
\vspace{-1.5em}
\vspaceSQ{-1em}
\label{fig:theory-mat}
\end{figure}

\section{THEORETICAL ANALYSIS}
\label{theory}

We conduct a theoretical analysis of the developed routing protocols using the
deployed SF network as a case study.
We focus on how well our routing uses the diversity of non-minimal paths,
which is necessary for high performance~\cite{besta2020fatpaths}.

\paragraph{Baselines and Parameters}
We analyze our layered routing that minimizes path overlap
(~\cref{sec:routing-des}) and compare it to a simple random layer
construction (RUES, Random Uniform Edge Selection) and to the 
state-of-the-art FatPaths scheme~\cite{besta2020fatpaths}.

We vary different parameters, including the fraction $p$ of preserved links in a layer, 
which refers to the proportion of links from the network that are included in each 
layer for the RUES scheme (specifically, we consider $p=40\%$, $p=60\%$, and $p=80\%$), 
and the number of layers used.
We focus on the deployed SF with 50 switches, but the results generalize
to larger sizes.
Overall, we show that the proposed layered routing is superior to the state-of-the-art 
in crucial metrics: lengths, distribution, and diversity of 
used paths, and the achieved throughput.

\subsection{Path Lengths}

The first important metric for evaluating routing is \emph{the length of paths
constructed using the proposed routing schemes}. 
Specifically, when routing in SF, one wants to use the single available
minimal path (with 1 or 2 hops, depending on picked switch pairs) and
the ``almost'' minimal ones -- with 3 hops -- as indicated in the FatPaths
study~\cite{besta2020fatpaths}.
To analyze whether the considered routing ensures this, we compute the average
and maximum lengths of the set of paths connecting each individual switch pair,
as produced by the respective routing schemes.
Fig.~\ref{fig:theory-pathlengths} shows the analysis results. 

Our novel layered scheme outperforms all others, because it ensures that
the highest fraction of switch pairs uses the ``almost'' minimal paths of
length \emph{at most} 3.
The downside of RUES is that the more randomness is employed, the larger the
maximum path length becomes. For a sampling factor $p=80\%$, there is no switch
pair with a path of length more than $4$, whereas for $p=40\%$ some switch pairs
have paths of length greater than $8$. This indicates large differences in path
lengths in different layers for some switch pairs, even if the average path
length is between $3$ and $4$. This can negatively impact load balancing
efforts as it becomes more difficult to predict path latency.
Then, in FatPaths, 
large fractions of switch pairs use 
paths of length 2, which means that these links may likely become congested.

Doubling the number of layers does not change the overall trends and it has
mostly no effect on the average path length distributions. Only the maximum
path lengths display a small shift to the right. This is because using more
layers increases the probability of finding a longer path.

\subsection{Path Distribution}

We now count the total number of paths that cross each
individual link, see Fig.~\ref{fig:theory-crossing}.
Our layered routing ensures a balanced scenario, i.e., 
close to equal utilization of each link. This
corresponds to a ``single bar'', i.e., the ``tighter''
the distribution the better balanced the paths are.

Similarly to the analysis on path length, less randomness leads to better
results, which is expected because as layers become less dense, the links that
are present will be more utilized. Hence, any link that by chance is included
in more than an average number of layers will have a higher number of crossing
paths and vice versa. FatPaths performs similarly to RUES for a sampling factor
of $p=80\%$. The distributions for $8$ layers
are slightly shifted to the right compared to 4 layers, as they have twice as
many paths.

\subsection{Path Diversity}
\label{sec::disj_paths}

Two paths are disjoint 
if they do not share common links. In layered routing, we aim to maximize
  the number of such paths used by node pairs.
Fig.~\ref{fig:theory-disjoint} displays counts of disjoint paths between
switch pairs. 
The FatPaths layer construction based on minimizing path overlap
underperforms because of its acyclic layers. 
Moreover, unlike in previous analyses, more randomness (and thus sparser
  layers) leads to better result for RUES. For a sampling factor of $p=40\%$ and
  $8$ layers, $\approx$97.5\% of switch pairs have at least the $3$ desired
  disjoint paths. This is the best performing algorithm out of the ones
  considered. However, this comes at the expense of disadvantageous path
  lengths and path distribution.

Our scheme does not need to make a similar trade-off
because with $8$ layers already around $88.5\%$ of switch pairs have at least
$3$ disjoint paths, which we have verified to grow to almost $100\%$ percent
when scaling to the next higher configuration that uses $16$ layers. At the
same time, the lengths and path distributions over links are highly beneficial.

\subsection{Maximum Achievable Throughput}

We also analyze the maximum achievable throughout (MAT).
MAT is defined as the maximum fraction of traffic demands from all endpoint pairs that can be accommodated simultaneously, 
while adhering to network and routing constraints. 
For example, a throughput of $1.5$ denotes that the network can sustain $1.5$ times the traffic demand of each communicating node pair simultaneously. 

Here, we consider an adversarial traffic pattern, which maximizes stress on the
interconnect by incorporating several large elephant flows between endpoints
that are separated by more than one inter-switch hop, and combining these large
flows with many small flows~\cite{jyothi2016measuring}.
We use TopoBench~\cite{jyothi2016measuring}, a throughput evaluation tool which
relies on linear programming to compute MAT. 
The results are displayed in Fig.~\ref{fig:theory-mat}.

\iftr
\else
\ifcr
Our algorithm outperforms FatPaths for different traffic intensities and layer
counts. This is most important for a small number of layers, which is key
for routing on IB hardware as using many layers reduces the supported network
sizes (cf.~Tab.~\ref{tab:uff}). Our layered routing experiences diminishing
returns beyond 16 layers. This is expected, as almost $100\%$ of endpoint pairs have at least $3$ disjoint paths for $16$ layers (one needs at least
that many disjoint paths to ensure high performance with non-minimal
routing). Before diminishing returns set in, FatPaths requires $8\times$ as many layers to reach equivalent performance, making our design much more practical.
\else
Our algorithm outperforms FatPaths for different traffic intensities and layer
counts. This is the most important for a small number of layers, which is key
for routing on IB hardware as using many layers decreases the supported network
sizes (see appendix Sec.~\ref{sec:max_net}). Our layered routing experiences diminishing
returns beyond 16 layers. This is expected, because almost $100\%$ of endpoint pairs have at least $3$ disjoint paths for $16$ layers (one needs at least
that many disjoint paths to ensure high performance with non-minimal
routing). Before diminishing returns set in, FatPaths requires $8\times$ as many layers to reach equivalent performance, making our design much more practical.
\fi
\fi

\subsection{Insights \& Takeaways - Theoretical Results}

Our novel IB layered routing achieves
superior performance in all considered path quality measures and especially in MAT. Almost around
$60\%$ of switch pairs have at least $3$ disjoint non-minimal paths when using
only $4$ layers, which grows to $88.5\%$ with $8$ layers.
Furthermore, we achieve the most balanced distribution of paths over the links in the
network. FatPaths performs similarly in terms of average and maximum path
lengths, but underperforms in the available number of disjoint paths per switch
pair. For RUES, a sampling factor of $p=60\%$ achieved the
most balanced results across all metrics, but RUES performs much worse in comparison
to FatPaths and our work overall.

\begin{table*}[t]
\caption{\textmd{Workload Configurations.}} 
\vspace{-0.5em}
\centering
\setlength{\tabcolsep}{4pt}
\scriptsize
\begin{tabular}{lllll}
\toprule
\textbf{Workload} & \textbf{Configuration} & \textbf{\# Nodes (N)} & \textbf{Scaling} & \textbf{Metric}\\
\midrule
Custom Alltoall & Message Sizes: $1$B $\to$ $4$MiB & $2,4,8,16,32,64,128,200$ & Weak & Bandwidth [MiB/s] \\
IMB Bcast/Allreduce~\cite{IMB-benchmarks} & Message Sizes: $1$B $\to$ $32$MiB & $2,4,8,16,32,64,128,200$ & Weak & Bandwidth [MiB/s] \\
eBB~\cite{hoefler2008switches} & Message Size: $128$MiB & $2,4,8,16,32,64,128,200$ & Strong & Bandwidth [MiB/s] \\
CoMD~\cite{comd}& $100^3$ Atoms per Process & $25, 50, 100, 200$ & Weak & Time [s] \\
FFVC~\cite{ffvc} &  $128^3$ Cuboid per Process for $\leq 64$ processes, else $64^3$ & $25, 50, 100, 200$ & Weak & Time [s] \\
mVMC~\cite{mvmc} & Unmodified \textit{job\_middle} weak-scaling test & $25, 50, 100, 200$ & Weak & Time [s] \\
MILC~\cite{milc-modeling,milc} & \textit{benchmark\_n8} Input & $25, 50, 100, 200$ & Weak & Time [s] \\
NTChem~\cite{ntchem} & \textit{taxol} Model & $25, 50, 100, 200$ & Strong & Time [s] \\
BFS$_{16}$~\cite{graph500, 7840705} & \# Vertices: $2^{23}$, $2^{24}$, $2^{25}$, $2^{26}$ Avg. Degree: $16$ & $25,50,100,200$ & Weak & Giga-Traversed Edges per Second [GTEPS] \\
BFS$_{128}$~\cite{graph500, 7840705} & \# Vertices: $2^{23}$, $2^{24}$, $2^{25}$, $2^{26}$ Avg. Degree: $128$ & $25,50,100,200$ & Weak & Giga-Traversed Edges per Second [GTEPS] \\
BFS$_{1024}$~\cite{graph500, 7840705} & \# Vertices: $2^{23}$, $2^{24}$, $2^{25}$, $2^{26}$ Avg. Degree: $1024$ & $25,50,100,200$ & Weak & Giga-Traversed Edges per Second [GTEPS] \\
HPL~\cite{hpl} & Matrix $A$ $\approx 1\text{ GiB }, 1\text{ GiB }, 1\text{ GiB } \text{ and }0.25 \text{ GiB }$ pre Process & $25, 50, 100, 200$ & Weak & Giga-Floating point OP/s [GFLOPS] \\
ResNet152~\cite{he2015deep, hoefler2022hamming} & Pure Data Parallelism & $40, 80, 120, 160, 200$ & Weak & Iteration Time [s] \\
Cosmoflow~\cite{hoefler2022hamming, mathuriya2018cosmoflow} & Model Shards: $4$ Data Shards: $\frac{\text{\# Nodes}}{4}$& $40, 80, 120, 160, 200$ & Weak & Iteration Time [s] \\
GPT-3~\cite{brown2020language, hoefler2022hamming} & Pipeline Stages (layers): $10$ Model Shards: $4$ Data Shards: $\frac{\text{\# Nodes}}{40}$ & $40, 80, 120, 160, 200$ & Weak & Iteration Time [s] \\
\bottomrule
\end{tabular}%
\label{tab:config}
\end{table*}

\begin{figure*}[t]
\vspaceSQ{-1em}
\vspace{-0.5em}
\centering
\vspaceSQ{-1em}
\begin{subfigure}[t]{0.24 \textwidth}
\captionsetup{justification=centering}
\centering
\includegraphics[width=\textwidth]{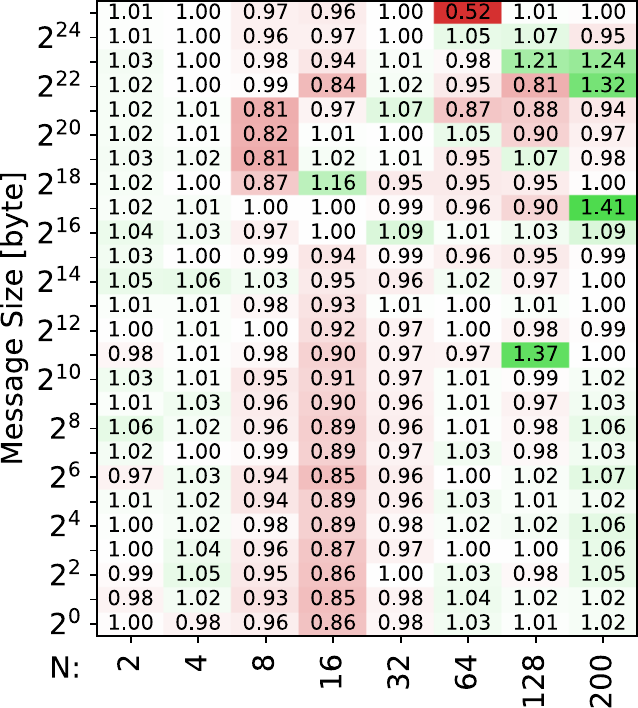}
\vspaceSQ{-1.5em}
\vspace{-1.5em}
\caption{\textmd{MPI Bcast - SF L vs. FT}}
\label{fig:speedup_imb_bcast_sf_l}
\end{subfigure}
\vspaceSQ{-1em}
\begin{subfigure}[t]{0.24 \textwidth}
\captionsetup{justification=centering}
\centering
\includegraphics[width=\textwidth]{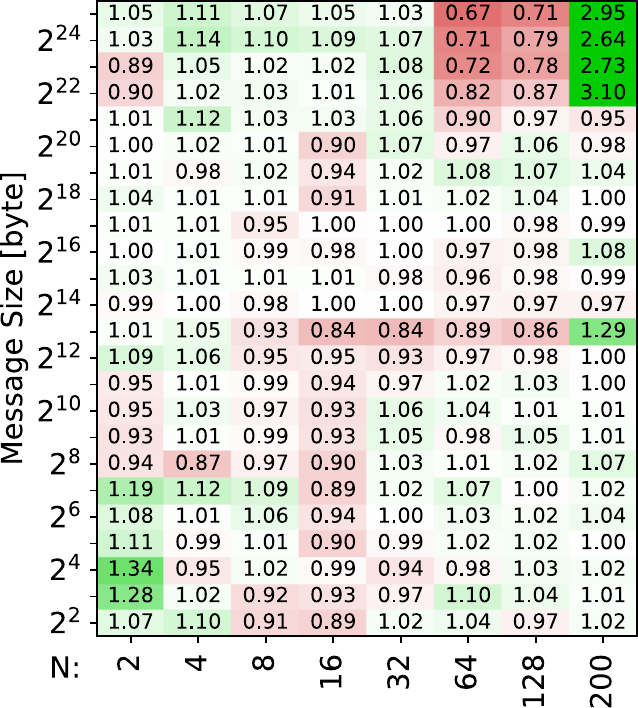}
\vspaceSQ{-1.5em}
\vspace{-1.5em}
\caption{\textmd{MPI Allreduce - SF L vs. FT}}
\label{fig:speedup_imb_allr_sf_l}
\end{subfigure}
\vspaceSQ{-1em}
\begin{subfigure}[t]{0.24 \textwidth}
\captionsetup{justification=centering}
\centering
\includegraphics[width=\textwidth]{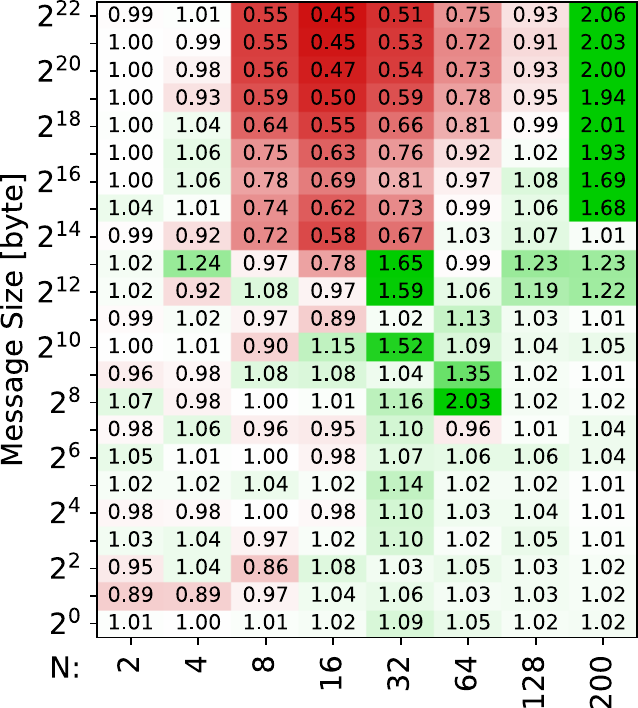}
\vspaceSQ{-1.5em}
\vspace{-1.5em}
\caption{\textmd{Custom Alltoall - SF L vs. FT}}
\label{fig:speedup_opt_a2a_sf_l}
\end{subfigure}
\vspaceSQ{-1em}
\vspaceSQ{-1em}
\begin{subfigure}[t]{0.24 \textwidth}
\captionsetup{justification=centering}
\centering
\includegraphics[width=1.0\textwidth]{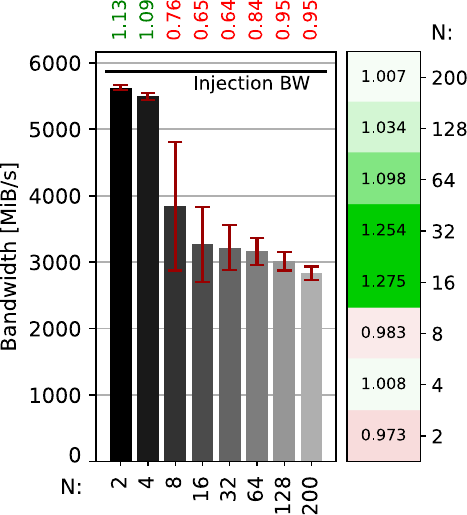}
\vspaceSQ{-1.5em}
\vspace{-1.5em}
\caption{\textmd{eBB - SF L vs. FT}}
\label{fig:ebb_sfl_cmb}
\end{subfigure}
\vspaceSQ{-1em}
\vspace{-0.5em}
\iftr
\else
\caption{\textmd{Relative performance difference of SF (linear placement strategy) over FT for various Microbenchmarks; eBB performance of SF L in comparison to maximum bandwidth and FT performance (higher is better), including routing improvement of this work over DFSSSP (heatmap).}}
\fi
\vspace{-1.5em}
\vspaceSQ{-1em}
\label{fig:microbenchmarks_add}
\end{figure*}

\ifcr
\begin{figure*}[t]
\vspaceSQ{-1em}
\centering
\vspaceSQ{-1em}
\begin{subfigure}[t]{0.24 \textwidth}
\captionsetup{justification=centering}
\centering
\includegraphics[width=\textwidth]{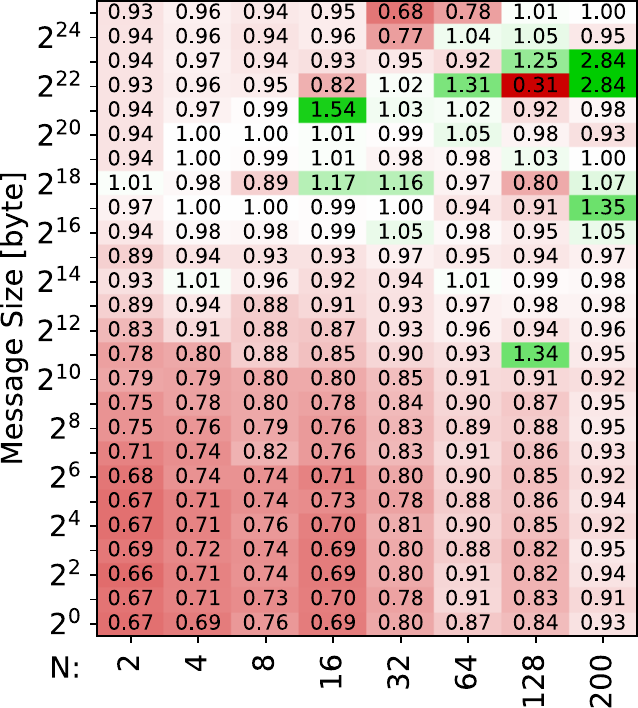}
\vspaceSQ{-1.5em}
\vspace{-1.5em}
\caption{\textmd{MPI Bcast - SF R vs. FT}}
\label{fig:speedup_imb_bcast_sf_r}
\end{subfigure}
\vspaceSQ{-1em}
\begin{subfigure}[t]{0.24 \textwidth}
\captionsetup{justification=centering}
\centering
\includegraphics[width=\textwidth]{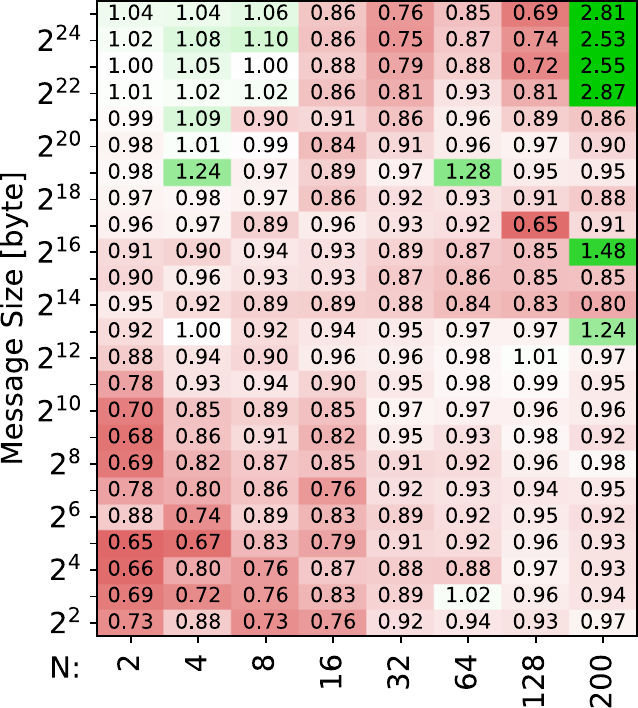}
\vspaceSQ{-1.5em}
\vspace{-1.5em}
\caption{\textmd{MPI Allreduce - SF R vs. FT}}
\label{fig:speedup_imb_allr_sf_r}
\end{subfigure}
\vspaceSQ{-1em}
\begin{subfigure}[t]{0.24 \textwidth}
\captionsetup{justification=centering}
\centering
\includegraphics[width=\textwidth]{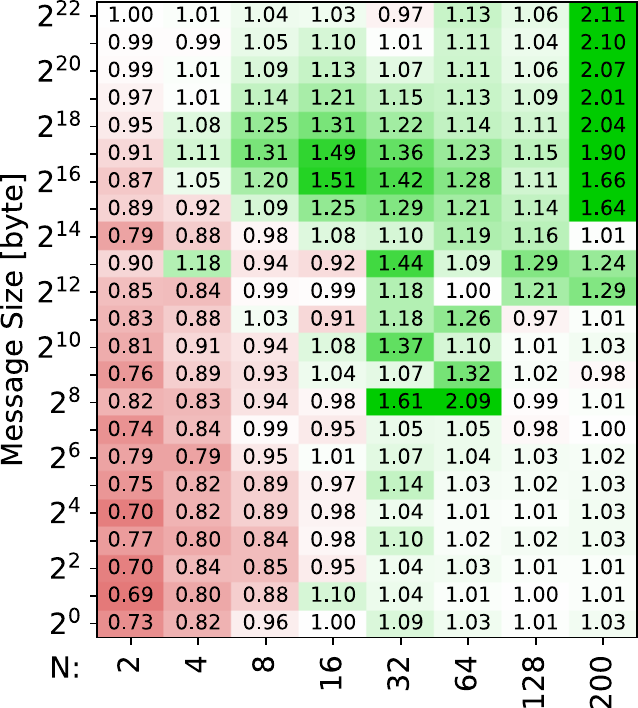}
\vspaceSQ{-1.5em}
\vspace{-1.5em}
\caption{\textmd{Custom Alltoall - SF R vs. FT}}
\label{fig:speedup_opt_a2a_sf_r}
\end{subfigure}
\vspaceSQ{-1em}
\vspaceSQ{-1em}
\begin{subfigure}[t]{0.24 \textwidth}
\captionsetup{justification=centering}
\centering
\includegraphics[width=1.0\textwidth]{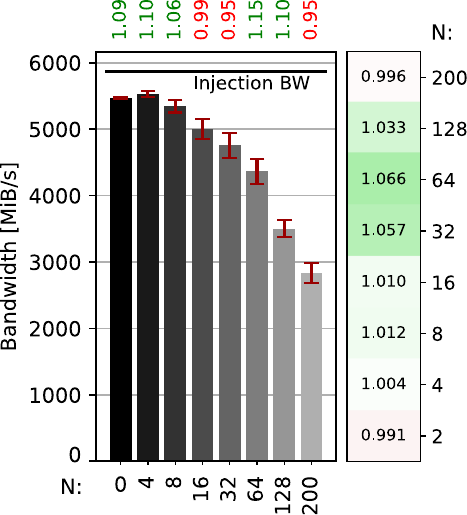}
\vspaceSQ{-1.5em}
\vspace{-1.5em}
\caption{\textmd{eBB - SF R vs. FT}}
\label{fig:ebb_sfr_cmb}
\end{subfigure}
\vspaceSQ{-1em}
\vspace{-0.5em}
\iftr
\caption{\textmd{Relative performance difference of SF (random placement strategy) over FT for various Microbenchmarks; eBB performance of SF R in comparison to maximum bandwidth and FT performance (higher is better), including routing improvement of this work over DFSSSP. Bidirectional bandwidth measured to be $5870\pm10 MiB/s$ using the ib\_*\_bw tools from the OFED \textit{perftest} package \cite{perftest}}}
\else
\caption{\textmd{Relative performance difference of SF (random placement strategy) over FT for various Microbenchmarks; eBB performance of SF R in comparison to maximum bandwidth and FT performance (higher is better), including routing improvement of this work over DFSSSP (heatmap).}}
\fi
\vspace{-1.5em}
\vspaceSQ{-1em}
\label{fig:mpi_collective_performance_a2a_allr}
\end{figure*}
\fi

\begin{figure}[t]
\vspaceSQ{-0.5em}
\centering
\includegraphics[width=0.73\linewidth]{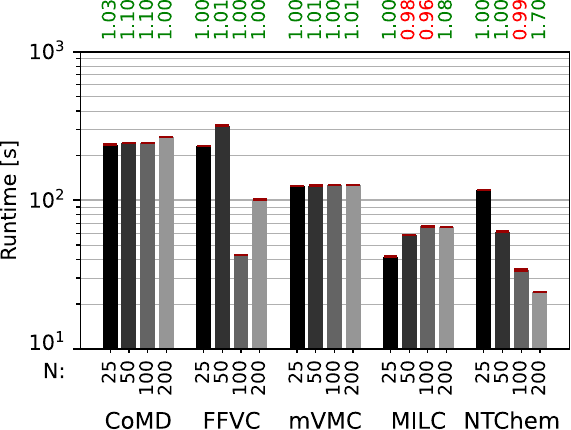}
\vspaceSQ{-0.5em}
\vspace{-0.5em}
\caption{\textmd{Runtime of scientific workloads (lower is better) - SF L vs. FT}}
\vspace{-0.7em}
\label{fig:sc_workloads_linear}
\end{figure}

\begin{figure}[t]
\vspaceSQ{-0.5em}
\centering
\includegraphics[width=0.65\linewidth]{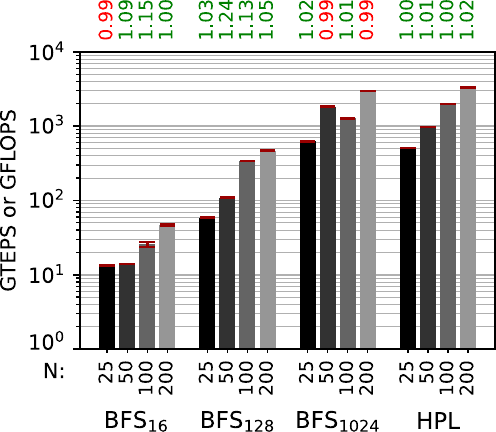}
\vspaceSQ{-0.5em}
\vspace{-0.5em}
\caption{\textmd{Performance of HPC benchmarks (higher is better) - SF L vs. FT}}
\vspace{-1.8em}
\label{fig:hpc_workloads_linear}
\end{figure}

\begin{figure}[t]
\vspace{0.0em}
\centering
\includegraphics[width=0.95\linewidth]{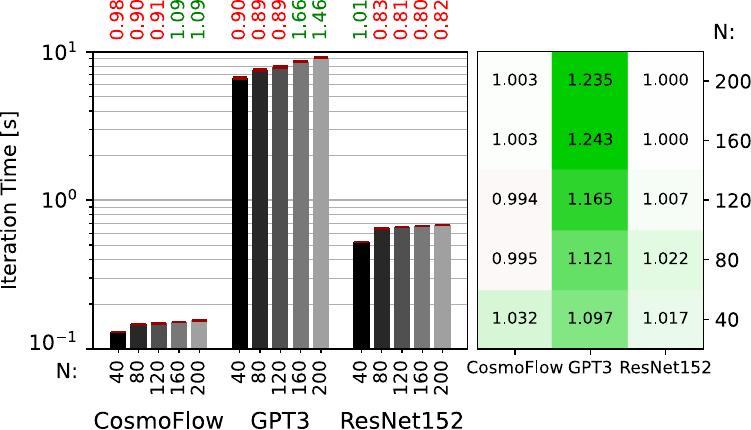}
\vspaceSQ{-0.5em}
\vspace{-0.5em}
\caption{\textmd{Iteration time of DNN proxy workloads (lower is better) SF L vs. FT and routing improvement of this work over DFSSSP (heatmap) for SF L.}}
\label{fig:dnn_workloads_linear}
\vspace{-1.8em}
\end{figure}

\section{EVALUATION}
\label{sec:eval}

We now illustrate the feasibility of our SF installation by evaluating
a broad set of applications from numerous domains against a comparable FT installation.

\subsection{2-Level Non-Blocking Fat-Tree}
FT topologies have historically been the usual choice for large-scale computing systems, 
largely due to their predictable behavior and full-bandwidth capabilities, when configured in a non-blocking manner. 
However, their high cost often leads to oversubscribed deployments at the tree's lowest level, 
reducing construction costs at the expense of bisection bandwidth.

To ensure a fair performance comparison with our SF installation, 
we construct a 2-level non-blocking FT, reusing the same hardware. 
The FT and SF both share the same network diameter and full-bandwidth capabilities. 
Our FT configuration employs 6 core and 12 leaf switches, compatible with our 36-port switches. 
Each leaf switch connects to each core switch through 3 links, and the remaining ports link to evenly distributed endpoints. 
This configuration supports up to 216 endpoints, making the FT marginally under-subscribed and thus strengthening the fairness of our comparison.

\subsection{Workloads \& Configurations}

We utilize a significant subset of the benchmarks included in the TSUBAME2 HyperX (t2hx) benchmark suite \cite{domke_hyperx_2019} and enhance them with a custom implementation of MPI\_Alltoall\footnote{Details on the performance improvements for the custom alltoall collective, over the default, can be found in the appendix (Sec~\ref{sec:a2a_improvement}).}, as well as three DNN proxies introduced by Hoefler et al. \cite{hoefler2022hamming}. 
The configuration of each benchmark is provided in Tab.~\ref{tab:config}.
Our analysis includes three classes of benchmarks:

\textbf{Microbenchmarks} 
We evaluate the system's bandwidth using Intel MPI Benchmarks' (IMB) measurements of the Allreduce and Bcast collectives~\cite{IMB-benchmarks},
and a custom alltoall.
We also assess 
the effective bisection bandwidth (ebb) of the system using Netgauge's eBB benchmark~\cite{hoefler2008switches}.

\textbf{Scientific Application \& HPC Benchmarks}
We evaluate a wide range of benchmarks, covering various scientific applications, all of which are listed in Tab.~\ref{tab:config} and taken directly from the t2hx benchmark suite.
We also analyze the performance of the High Performance Linpack (HPL)~\cite{hpl} benchmark and of the breadth-first search (BFS)~\cite{besta2017slimsell} in the Graph 500 Benchmark \cite{graph500}. 
Additionally, we extend the BFS performance analysis by changing the average degree of the vertices (edgefactor), 
while scaling the number of vertices linearly with the number of participating compute nodes. 
Specifically, we consider edgefactors $16$, $128$ and $1024$.

\textbf{DNN Proxies} 
The DNN proxies evaluated on SF include ResNet152~\cite{he2015deep} (pure data parallelism), CosmoFlow~\cite{mathuriya2018cosmoflow} (data and operator parallelism) and GPT-3~\cite{brown2020language} (data, operator, and pipeline parallelism), as outlined in Tab.~\ref{tab:config}. For GPT-3, each pipeline stage processes one DNN-layer.

\subsection{Execution Environment}
To ensure consistency and reproducibility, all benchmarks were compiled using GCC v4.8.5 and executed using OpenMPI v1.10.7.
We use one MPI rank per node and assign one OpenMP thread per physical core on Socket 1 of the dual-socket system
(pinning on Socket 2 introduces non-negligible slowdowns due to inter-socket communication).

We investigate two MPI rank placement strategies: linear and random.
The linear strategy places rank $j$ on node $j$, a commonly used approach that enhances latency and traffic locality, especially for FTs ~\cite{michelogiannakis2017aphid, slurm132}.
This strategy also models a system with minimal fragmentation.
In contrast, the random strategy represents systems with significant fragmentation. It randomizes rank placement to potentially reduce network bottlenecks on SF, albeit at the cost of increased latency.
For FT, the linear placement significantly outperformed its random counterpart in all microbenchmarks and exhibited comparable performance in the remaining tests. Consequently, we report SF performance relative to the FT's linear placement only.

Each benchmark configuration is executed five times; microbenchmarks are executed for at least 100 iterations.
We assess all SF benchmarks using our new multipath routing algorithm based on both minimal and almost minimal paths, 
as well as the defacto standard multipath routing algorithm in IB (DFSSSP), that leverages minimal paths only~\cite{domke-hoefler-dfsssp}. 
We instantiate each routing algorithm once with 1, 2, 4, and 8 layers, respectively, but only report the results of the best-performing variant for each benchmark configuration. 
For all FT benchmarks we choose the commonly used ftree routing~\cite{Jacobs2010DModKRP}.
Mean and standard deviation of the results are reported, with the latter indicated using red error bars for all bar plots.
Relative performance differences of SF over FT are annotated above each bar. 
Any significant performance gains or losses of our novel routing algorithm in comparison to DFSSSP for any benchmark are
either explicitly stated in the text or visualized using heatmaps.

\ifcr
In the main text, we present comprehensive results for SF using the linear placement strategy, and include only microbenchmark results for the random placement strategy due to space considerations. 
Detailed results of the random strategy for other benchmarks, which largely mirror those obtained with the linear strategy, are in~\cref{sec:add_res}.
\fi

\subsection{Microbenchmarks}
\ifcr
Fig.~\ref{fig:speedup_imb_bcast_sf_l}--\ref{fig:speedup_opt_a2a_sf_l} illustrate the relative performance differences of SF with linear placement over FT and Fig.~\ref{fig:speedup_imb_bcast_sf_r}--\ref{fig:speedup_opt_a2a_sf_r} of SF with random placement over FT for MPI collectives bcast, allreduce, and custom alltoall. 

Generally, SF's performance using the linear placement strategy closely matches that of the FT, with FT only displaying minor advantages in bcast and allreduce for 8 and 16 node configurations at smaller, latency-sensitive message sizes. 
This marginal edge of FT in specific configurations is due to its architecture, wherein leaf switches connect to at least 16 nodes, facilitating localized communication with zero inter-switch hops, thus minimizing latency.
While SF, under linear placement, enjoys the benefits of zero inter-switch hops mostly for configurations of up to 4 nodes -- owing to its design of connecting 4 nodes per switch -- random placement generally does not benefit from this localized communication advantage. 
As a result, SF experiences marginally lower performance in comparison to FT for these latency-sensitive scenarios with the random placement strategy.

In contrast, for the communication-intensive alltoall collective, SF's performance closely mirrors, or even slightly surpasses, that of the FT for small message sizes when employing the linear placement strategy (cf. Fig.~\ref{fig:speedup_opt_a2a_sf_l}). 
However, in 8, 16, and 32 node configurations, particularly with bandwidth-critical message sizes, SF lags due to congestion caused by all inter-switch communication occurring between 2, 4, or 8 switches, respectively. 
This leads to traffic bottlenecks on the often single shortest path between these switches. 
While our new routing scheme, as discussed in~\cref{theory}, theoretically mitigates this congestion, the absence of adaptive load balancing limits practical improvements to at most $7\%$ over DFSSSP.

Switching to the random placement strategy markedly improves SF's performance for the alltoall collective, as shown in Fig.~\ref{fig:speedup_opt_a2a_sf_r}. 
This strategy not only overcomes the noted bottlenecks but also enables SF to significantly outperform FT. 
This improvement results from the random placement strategy's enhanced traffic distribution across the network, showcasing the trade-off between increased latency for smaller message sizes and superior traffic balancing within the SF topology. 
These findings imply that the integration of adaptive load balancing with our routing scheme could effectively address the congestion issues identified with linear placement, underscoring the potential of our routing scheme to optimize network performance for demanding communication patterns.

Lastly, in Fig.~\ref{fig:ebb_sfl_cmb} and Fig.~\ref{fig:ebb_sfr_cmb}, we present the ebb across various node counts for the linear and random placement strategy, respectively.
At maximum node count we achieve approximately half of the injection bandwidth, equating to $75\%$ of the theoretical bisection bandwidth optimum~\cite{besta2014slim}, with both strategies.
Though the FT matches SF's full-system ebb, it outperforms SF with linear placement for the 8, 16, and 32 nodes configurations.
This discrepancy mirrors the observations for the alltoall collective and is similarly overcome with the random placement strategy (cf. Fig.~\ref{fig:ebb_sfr_cmb}).

In the right section of both Fig.~\ref{fig:ebb_sfl_cmb} and Fig.~\ref{fig:ebb_sfr_cmb}, heatmaps display the performance gains of our new routing scheme over DFSSSP for the eBB benchmark. 
Notably, for the linear placement strategy, improvements of up to $28\%$ are observed for the earlier described node configurations, which are especially prone to congestion.
Under the random placement strategy, the level of improvement is less significant, with only up to $7\%$, suggesting that this strategy's primary advantage lies in its ability to distribute traffic more evenly, even in the absence of adaptive load balancing. 
\fi

\subsection{Scientific Workloads \& HPC Benchmarks}
In Fig.~\ref{fig:sc_workloads_linear}, we present the runtime and relative performance of the solver/kernel for each of the scientific workloads on SF, using the linear placement strategy.
The scaling behavior of each workload, based on their configurations detailed in Tab.~\ref{tab:config}, is evident.
Notably, the drop in runtime for FFVC when scaling from 50 to 100 nodes is due to the decrease in the workload's problem size when running on $>64$ nodes.
Utilizing almost minimal paths in combination with minimal paths does not generate any significant speedup for these workloads over pure minimal routing (DFSSSP), 
and generally results in only small runtime variances of $<1\%$. This is due to the communication time only constituting a small fraction of the overall runtime for these scientific workloads~\cite{domke_hyperx_2019, 10.1007/978-3-319-58667-0_12}. 

Fig.~\ref{fig:hpc_workloads_linear} shows the performance of the HPC benchmarks, 
which display similar weak-scaling behavior as the scientific workloads. 
HPL exhibits almost linear scaling performance when increasing the number of nodes from $25$ to $50$ or $100$ nodes, 
indicating that the overhead introduced by the increased amount of communication is negligible. 
Consistent with these results, introducing almost minimal paths to the routing impacts performance by less than $1\%$ for the HPL benchmark.
The only exception is the $200$ node setting, where the decrease in the problem size (per node) is likely the main cause for the deviation from the linear scaling observed.

In the case of the Graph 500 - BFS benchmark, we experienced high variance with the default implementation. To mitigate this, we fixed the seed for the graph generation and used the same source vertex for each BFS run. The BFS scaling results show more fluctuations in comparison to the HPL results, particularly for the sparser variant. This is accompanied by greater variability in speedup through almost minimal paths, which ranged from -$5\%$ to +$1\%$. It is not clear whether this can be attributed purely to network communication or to other factors such as caching effects and system noise.

Overall, our experiments show SF competes effectively with FT in terms of performance, while being very effective for scaling both scientific workloads and HPC benchmarks, even when limited to minimal paths. 
\subsection{Deep Learning Workloads}

The left part of Fig.~\ref{fig:dnn_workloads_linear} shows the runtime and relative performance of the DNN proxies when linearly increasing the number of nodes from $40$ to $200$. ResNet152 with pure data parallelism only requires allreduce for gradient aggregation. CosmoFlow with a hybrid of data and operator parallelism requires allgather, reduce-scatter, allreduce, and point-to-point communications. GPT-3 with a hybrid of data, operator, and pipeline parallelism requires allreduce and point-to-point communications. As we increase the data shards proportionally to the number of nodes, the scalability is mainly determined by allreduce across the data dimension. 

We find that CosmoFlow's runtime on SF is comparable to that on FT. 
In contrast, GPT-3 notably performs better on SF for configurations with 160 and 200 nodes, while ResNet152 begins to lag as the node count increases. 
Although both GPT-3 and ResNet152 predominantly rely on allreduces at higher node counts, their diverging performance trends can be attributed to differences in message sizes; GPT-3 handles significantly larger messages than ResNet152. 
Expectedly, the performance trend of GPT-3 matches the trend of MPI Allreduce for the high node count configurations (cf. Fig.~\ref{fig:speedup_imb_allr_sf_l}).

The right part of Fig.~\ref{fig:dnn_workloads_linear} shows that our work generally outperforms DFSSSP for GPT-3, with up to $24\%$ improvements. 

\subsection{Insights \& Takeaways - Empirical Results}
When analyzing communication-intensive workloads on configurations with 8, 16, or 32 nodes, we identified some congestion challenges. These challenges stemmed from the non-adaptive nature of the path selection. However, by employing a random placement strategy, these issues were effectively counteracted.
Our findings subsequently indicate that SF consistently achieves performance on par with, or even surpassing, the well-established FT topology, particularly under conditions of full-system utilization.
Additionally, SF displays effective scaling capabilities across a diverse range of workloads.
In comparison to the established DFSSSP, our novel routing approach exhibited promising performance, registering improvements of over $20\%$.

\subsection{Scalability \& Cost Analysis}
\label{sec:scalability_cost}

\begin{table*}[t]
\caption{\textmd{Maximal scalability and costs of SF deployments compared to non-blocking FT2, FT2 oversubscribed by 3 (FT2-B), FT3 and 2-D HyperX (HX2) under given port constraints. For the fixed size cluster we use 64-port switches for the FT2 and FT-B, 40-port switch for HX2, and 36-port for SF and FT3.}} 
\vspace{-0.6em}
\centering
\setlength{\tabcolsep}{3pt}
\scriptsize
\begin{tabular}{l|ccccc|ccccc|ccccc|ccccc}
\toprule
    \multicolumn{1}{c}{} & \multicolumn{5}{c}{\textbf{36-port switches}} & \multicolumn{5}{c}{\textbf{40-port switches}} & \multicolumn{5}{c}{\textbf{64-port switches}} & \multicolumn{5}{c}{\textbf{2048 nodes clusters}} \\
    \textbf{}   & FT2  & FT2-B & FT3 &    HX2     & SF    & FT2 & FT2-B  & FT3 & HX2   &  SF   & FT2 & FT2-B   & FT3 & HX2 &  SF & FT2 & FT2-B & FT3 & HX2 &  SF\\ \midrule
    Endpoints   & 648 & 972  &  11664  &   2028    & 6144  & 800 & 1200  & 16000  & 2744 & 7514  & 2048  & 3072 & 65536  & 10648  &  32928 & 2048 & 2048 & 2048 & 2197 & 2178 \\
    Switches    & 54   & 45 & 1620    &   169     & 512   & 60   & 50 & 2000    & 196   & 578   & 96   & 80 & 5120    &  484  &   1568 & 96  & 59 & 303   & 169 & 242 \\
    Links    & 648   & 324 & 23328   &  2028     & 6144  & 800  & 400 & 32000   & 2548  &  7225     & 2048 & 1024 & 131072  & 10164   &  32928 & 2048 & 344 & 4320  & 2028 & 2057\\ \midrule
    Costs [M\$]     & 1.5 & 1.1 &   45   &  4.5  & 13.8 & 2.4 & 1.7 & 84.2 & 7.8  & 22.4 & 9 & 7.2 & 491 & 45.5 & 146 & 7.5 & 2.7 & 8.3 & 6.4 & 5.8\\
    Cost/Endp [k\$]  & 2.2 & 1.2 &  3.8 &  2.2  & 2.2 & 3 & 1.5 & 5.2 & 2.8 & 2.9 & 4.4 & 2.3 & 7.5 & 4.3 & 4.4 & 3.6 & 1.3 & 4 & 3.1 & 2.8  \\

\bottomrule
\end{tabular}%
\label{tab:costs}
\vspace{-1.5em}
\end{table*}

FT topologies are the preferred choice for large-scale HPC deployments due to their adaptability, adoptable bisection bandwidth, established routing, and isolation advantages. These properties often benefit application performance consistency ~\cite{stunkel2020high,varrette2022aggregating,bhatele2019analyzing}. However, their low-diameter configurations do not scale as well as contemporary topologies~\cite{kathareios2015cost}.

We compare the scalability and deployment cost of 2-level FTs (FT2), 3-level FTs (FT3), 2-D HyperX (HX2) \cite{domke_hyperx_2019, Ahn:2009:HTR:1654059.1654101}, and SF. 
Our evaluation, summarized in Tab.~\ref{tab:costs}, includes both the non-blocking FT2 variant and its 3:1 oversubscribed version (FT2-B). 
The pricing details are in~\cref{sec:costs}.

\paragraph{Scalability} 
We show that SF networks offer a distinct advantage in scalability by evaluating maximum network size for a HW setup with 36, 40, and 64-port switches. SF can accommodate approximately 10, 6, and 3 times more endpoints than FT2, FT2-B, and HX2 respectively, while maintaining a lower or comparable cost-to-endpoint ratio and the same network diameter of 2. FT3 can accommodate more endpoints than SF, however, this comes at a significantly larger (around 1.75x) cost-to-endpoint ratio and increased network diameter which has an impact on a performance of latency critical applications.
This makes SF a compelling choice for large-scale diameter-2 deployments.

\paragraph{Cost} 
When the number of endpoints is predetermined, SF's requirement for fewer port switches can reduce overall deployment costs, while keeping comparable benchmark performance to FT2 as shown in~\cref{sec:eval}. Tab.~\ref{tab:costs} further shows an example of fixing a cluster requirement to 2048 endpoints. Realising such a cluster using SF in comparison to FT2, HX2, and FT3 results in absolute cost saving of \$1.7M, \$0.6M, and \$2.5M respectively. While using FT2-B might be cheaper in this scenario, it does not provide the full bandwidth property as SF, FT2, HX2, and FT3.

\section{RELATED WORK}

Our work touches on different areas. We now outline related works, excluding
the ones covered in past sections.

\paragraph{Network Topologies}
Several recent networks build upon SF. This includes
Megafly~\cite{flajslik2018megafly}, Bundlefly~\cite{bundlefly_2020},
Galaxyfly~\cite{lei2016galaxyfly}, and Xpander~\cite{valadarsky2015}. 
Yet, they do not provide diameter-2 and thus none of them are competitive with
SF in latency, cost, or power consumption, as observed by recent
results~\cite{besta2020fatpaths}. Although PolarFly has shown promising results 
in recent studies, its advantages over SF can be attributed to the diligent 
design of routing protocols that leverage its structure to guarantee 
optimal routing decisions \cite{lakhotia2022polarfly, lakhotia2023innetwork}.
Some recent designs based on similar principles target on-chip networks only~\cite{besta2018slim, iff2022sparse}.

\paragraph{Physical Interconnect Installations}
The majority of works on interconnects use
simulations for evaluation~\cite{besta2014slim, dally08, dally07, 
valadarsky2015, flajslik2018megafly, bundlefly_2020, lei2016galaxyfly, singla2012jellyfish, ahn2009hyperx,
DBLP:conf/isca/KoibuchiMAHC12, besta2021towards}.
However, some topologies have been evaluated with real installations.
This includes -- for example -- HyperX~\cite{domke_hyperx_2019}
and Dragonfly~\cite{aries}.
Here, we offer the first real evaluation of Slim Fly.

\paragraph{Congestion Control \& Load Balancing}
In general, we do not focus on transport protocols (flow, congestion).  Here,
we rely on mechanisms from the FatPaths~\cite{besta2020fatpaths} architecture.
In layered routing, traffic is balanced across layers. We use simple
randomized and round-robin schemes, which results in high performance.
Other schemes could also be incorporated, including load
balancing based on flows~\cite{hopps2000analysis, curtis2011mahout, rasley2014planck,
sen2013localflow, tso2013longer, benson2011microte, zhou2014wcmp, al2010hedera,
kabbani2014flowbender}, flowcells~\cite{he2015presto},
flowlets~\cite{katta2016clove, alizadeh2014conga, vanini2017letflow,
katta2016hula, kandula2007dynamic}, and single packets~\cite{zats2012detail,
handley2017re, dixit2013impact, cao2013per, perry2015fastpass, raiciu2011improving}.%

\section{CONCLUSION}

Slim Fly (SF) is the first network topology that lowered cost and improved performance by reducing the network diameter to two, promising significant improvement over established interconnects.
However, it has not yet been tested in practice.
We address this by deploying the first at-scale SF installation and establishing and implementing open-source routines for cabling and physical layout, to guide future deployments and effectively verify cabling.
This can foster the adoption of SFs in broad industry and facilitate practical deployments of other low-diameter topologies, including the most recent ones, such as PolarFly or Bundlefly.

We further introduce a novel high-performance routing scheme that improves upon state of the art, 
achieving up to $24\%$ speedup for the evaluated deep neural network (DNN) workloads over the standard IB multipath routing algorithm (DFSSSP) through non-minimal paths.

We use the first practical, real-world deployment of SF to demonstrate the topology's ability to scalably process a wide selection of modern workloads such as distributed DNN training, graph analytics, or linear algebra kernels.
It consistently matches or surpasses the performance of a comparable non-blocking Fat Tree (FT) deployment for a wide selection of workloads, for example, achieving a 66\% speedup for distributed deep neural network training. Importantly, SF simultaneously delivers superior scalability.
For example, it enables connecting between $3\times$ and $10\times$ the number of servers compared to other diameter-2 topologies like 2-level FT and 2-D HyperX, while maintaining both a comparable cost-to-endpoint ratio and full bandwidth. 
For larger installation sizes, SF's scalability translates to significant cost advantages, for example, 50\% over full bandwidth non-blocking 3-level Fat Tree configurations~\cite{besta2014slim}.
Overall, this effort will spearhead future research into more powerful network topologies.

\ifbd
\else
\section*{Acknowledgments}
We thank Colin McMurtrie, Mark Klein, Angelo Mangili, and the whole CSCS team granting access to the Ault and Daint machines, and for their excellent technical support with the Slim Fly cluster infrastructure. 
We thank Timo Schneider for help with infrastructure at SPCL.
This project received funding from the European Research Council (Project PSAP, No.~101002047), and the European High-Performance Computing Joint Undertaking (JU) under grant agreement No.~955513 (MAELSTROM). 
This project received funding from the European Union’s HE research and innovation programme under the grant agreement No. 101070141 (Project GLACIATION).
This project was supported by JSPS KAKENHI Grant Number JP19H04119.

\fi

\printbibliography

\appendix
\section{Details of Slim Fly Construction}
\label{sec:sf_cons}

\subsection{Selecting Topology Size, Parametrizing Input}
\
Overall, one first chooses a prime power $q$ that satisfies the equation $q =
4w + \delta$ for some $\delta \in \{-1, 0, 1\}$ and $w \in \mathbb{N}$. $q$ is
an input parameter that determines the whole topology structure. For example,
the number of vertices (switches) is $N_r = 2 q^2$ and the network radix $k' =
\frac{3q-\delta}{2}$.
In our case, $N_r = 50$, thus $q = 5$, which satisfies the equation $q = 4w +
\delta$ for $w = 1$, $\delta = 1$, and $k' = 7$. Hence, every switch is
connected to $7$ other switches.
Interestingly, this construction forms the famous Hoffman-Singleton
graph~\cite{hoffman1960moore, hafner2003hoffman}, which is \emph{optimal} with
respect to the Moore Bound.
Finally, as a regular and direct network, it is recommended to attach $p =
\ceil*{\frac{k'}{2}}$ endpoints to each switch to ensure \emph{full global
bandwidth}~\cite{besta2014slim}. In our case, $p = 4$.

\subsection{Finding Needed Algebraic Structures}
\
Once $q$ is selected, one uses it to construct several algebraic structures.
Specifically, one finds a \emph{base ring}~$\mathbb{Z}_q$ (for us,
$\mathbb{Z}_5 = \{0, 1, ..., 4\}$), its \emph{primitive element}~$\xi$ that
generates all elements of $\mathbb{Z}_q$ (for us, $\xi = 2$), and two
\emph{generator sets} $X = \{\xi^0, \xi^2, ..., \xi^{q-3}\}$ and $X' = \{\xi^1,
\xi^3, ..., \xi^{q-2}\}$ (for our installation, $X = \{1,4\}$ and $X' =
\{2,3\}$).
While not complex, details on these structures are not necessary to understand
our Slim Fly deployment. The interested readers may check them in the original 
publication~\cite{besta2014slim}.

\subsection{Labeling and Connecting Switches}
\label{sec:sf_cons_eq}
\
Each switch receives a 3-tuple label from a set $\{0,1\} \times \mathbb{Z}_q
\times \mathbb{Z}_q$.
Thus, SF switches come in two flavors determined by the first elements of their
labels: $(0, \cdot, \cdot)$ and $(1, \cdot, \cdot)$.
These labels determine how the switches are connected. Specifically, switches
with labels $(0, \cdot, \cdot)$ are connected using the following
equation~\cite{besta2014slim}:

\small
\begin{equation}\label{equation::first}
    \text{switch }(0,x,y)\text{ is connected to }(0,x,y') \iff y-y' \in X.
\end{equation}
\normalsize

Symmetrically, switches with labels $(1, \cdot, \cdot)$ use the following
equation:

\small
\begin{equation}\label{equation::second}
    \text{switch }(1,m,c)\text{ is connected to }(1,m,c') \iff c-c' \in X'.
\end{equation}
\normalsize

Lastly, two switches with labels $(0, \cdot, \cdot)$ and $(1, \cdot, \cdot)$,
respectively, are connected according to the following equation: 

\small
\begin{equation}\label{equation::third}
    \text{switch }(0,x,y)\text{ is connected to }(1,m,c) \iff y=m\cdot x + c
\end{equation}
\normalsize

\subsection{Topology Structure \& Physical Layout}
\
The graph underlying the SF topology consists of two same-size subgraphs.
One subgraph contains routers $(0, x, y)$, the other consists of routers $(1,
m, c)$. Each subgraph contains~$q$ identical groups of routers. Groups in
different subgraphs usually differ from one another.  There are no connections
between groups within the same subgraph, i.e., no two routers $(0, x, y)$ from
different groups are linked, the same holds for routers $(1, m, c)$. However, each
group from one subgraph has connections to \emph{every other group} in the
other subgraph; thus the groups form a fully connected bipartite graph.  

This property facilitates physical layout and we use it in our installation.
Specifically, as recommended in the original work~\cite{besta2014slim}, we
combine groups from different subgraphs pairwise; these combined groups form
racks. In general, this leads to $q$ racks, each with $2q$ routers. In our
installation, we have 5 racks, each with 10 routers and 40 compute nodes.

\subsection{Constructing Slim Fly with N nodes}
As the space of valid SF topologies is quite sparse, we show the simple steps needed to find
a SF network with the number of nodes as close to N as possible:
\begin{enumerate}
    \item Obtain the cube root R of the desired node count N
    \item Find prime powers close to R
    \item Obtain the corresponding full-bandwidth network configurations (see previous sections)
    \item Verify network sizes and select the network that is closest to N in terms of number of supported nodes
\end{enumerate}

\section{Routing Details}
\label{sec:routing_details}

\subsection{Details of Layer Generation}
\label{sec:layer_gen}

We provide more details on crucial parts of layer generation.

\subsubsection{Finding Almost-Minimal Paths}
\label{amp}
\
We look for almost-minimal paths that are exactly 3 hops long (one hop longer
than SF's diameter of two), while balancing the number of paths crossing each link 
(to avoid highly congested links). We do not target longer paths, in order to conserve
network resources (i.e., a flow taking fewer hops occupies fewer buffers).

For this, we design a heuristic based on a modified breadth first search graph
traversal starting from the source node $src$, constraining the path length to
3. In theory one could also define a range of valid lengths.  
The heuristic obtains the set $P$ of all valid paths
starting in $src$ and ending in the destination node $dst$; $P = \{(u_1, \dots,
u_l) \mid l = 3 \wedge u_1 = src \wedge u_l = dst\}$. Here, a path is
considered valid if it satisfies the given length constraint $3$ and if
its insertion into the layer does not affect any previously inserted paths.
Then, we choose a path $p \in P$ that minimizes link weights, i.e., $\forall p'
\in P\  \omega(p) \le \omega(p')$ where $\omega(p)$ is the sum of weights of
links included in $p$.

\subsubsection{Node Pair Priority Queue}
\ 
The order in which the paths are inserted is very important, because it may
impact whether we are able to find new paths. If one would first find a given
number of paths for a single node pair, and only then proceed to the next node
pair, some node pairs might not receive any, or much fewer, paths than other
pairs.  To alleviate this, we balance the total number of added almost-minimal
paths across all node pairs.  For this, each node pair is assigned a priority
value, equal to its total number of almost-minimal paths across all layers; the
lower the value, the more important it is to find a path for this node pair.
Therefore, the number of required priority levels is upper-bounded by $|L|-1$,
because each node pair can have at most one almost-minimal path per each of
$|L|-1$ layers, and is initially in the highest priority value (value of $0$).
The lowest priority level is value $|L|-1$, which only contains node pairs who
have had an almost-minimal path inserted in every layer.

Whenever a path is added to a layer, all of the node pairs that have a
non-minimal path inserted have their priority decreased and they move up
to the next higher priority layer. For instance, in Fig.~\ref{fig:path} by
assuming that the minimum length for an almost-minimal path is two, adding the
illustrated path to a layer, results in both node pairs $(v_1, v_4)$ and $(v_2,
v_4)$ having an almost-minimal path added for them 
(assuming we allow paths of length $2$ and $3$ as non-minimal paths and $dst$ is 
one of the receiving nodes). Therefore, both of their
priorities would be decreased by $1$. This also assumes that the paths were not
already in this layer, which could have been the case for $(v_2, v_4)$.

The $node\_pairs$ list generated from the priority queue $p$ in
Algorithm~\ref{algo::pseudo_layer} contains the entries of the priority queue in
the order of priority value, and randomized within each level. Hence, the layer
generation algorithm first tries to add an almost-minimal path for all nodes of
priority value~$0$ in a random order, and then move to the nodes of the next
value. Hence, it first processes all node pairs with no inserted paths, then
with one inserted path, and so forth, facilitating a balanced path distribution
across node pairs.

\subsubsection{Path Weighting}
\begin{figure}[h]
    \centering
    \includegraphics[width=\columnwidth]{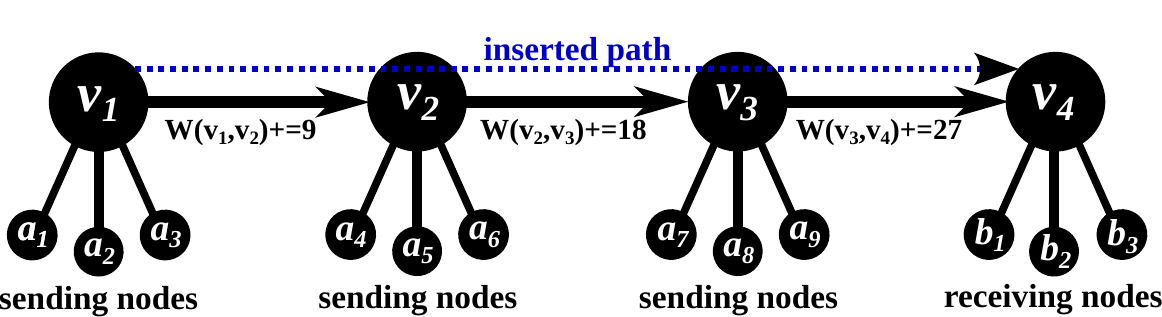}
    \caption{\textmd{Illustration of the weight update methodology employed by the
algorithm. After the insertion of the path from $v_1$ to $v_4$, the weights of
the links $(v_1, v_2)$, $(v_2, v_3)$ and $(v_3, v_4)$ are increased by $9$,
$18$, and $27$, respectively.}} \label{fig::weight_update}
\end{figure}

\ 
A weight update is performed after the insertion of a new path into any layer.
The weight of each link in any existing path is increased by the total number
of new ``routes'' that now occupy the link.  An example is shown in
Fig.~\ref{fig::weight_update}. The weight of link $(v_1, v_2)$ is increased
by $9$ because it has $9$ new routes using it, as there are $3$ sending nodes
($a_1-a_3$) and $3$ receiving nodes ($b_1-b_3$).  The weight of link $(v_3,
v_4)$ is increased by $27$ as there are $9$ sending nodes ($a_1-a_9$) and $3$
receiving nodes ($b_1-b_3$). 

\subsubsection{Potential Invalidity of Paths} \label{sec:invalids}
\begin{figure}[h]
    \centering
    \includegraphics[width=\columnwidth]{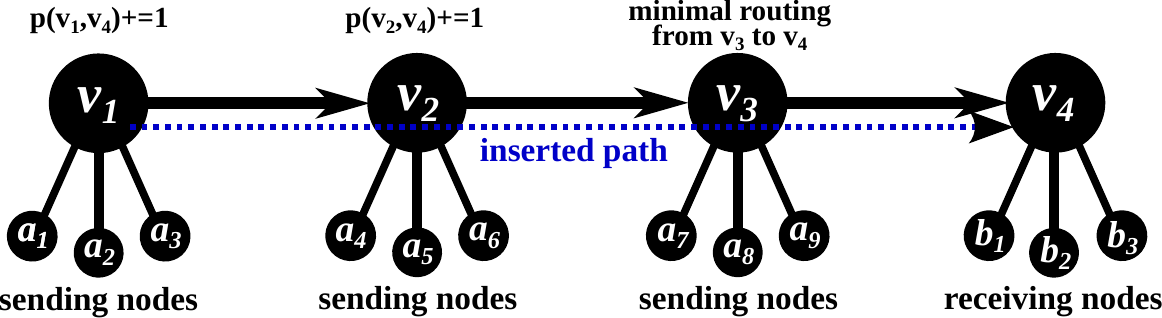}
    \caption{\textmd{Illustration of an almost-minimal path from $v_1$ to $v_4$, which
    enforces minimal routing from $src$ nodes like $a_7$, located on the sub-paths, to $dst$ nodes, i.e. $b_1$, in this layer.}}
    \label{fig:path}
\end{figure}

\begin{figure}[h]
    \centering
    \includegraphics[width=\columnwidth]{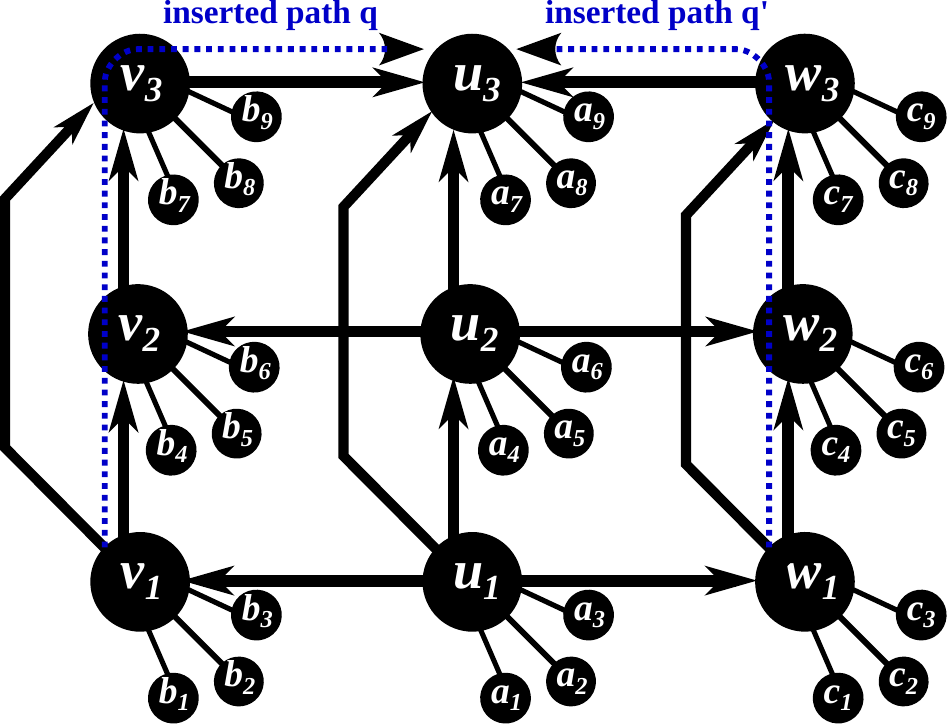}
\caption{\textmd{Illustration of a scenario in which no almost-minimal, valid path of
length exactly $3$ can be found for node pair $(u_1, u_3)$ in the given layer
due to the prior insertion of two valid paths.}}
    \label{fig:paths}
\end{figure}

\ 
For a given source $src$ and destination $dst$, it may happen that $P =
\emptyset$, in which case no almost-minimal path is added to a given layer for
that node pair. 
There are two scenarios when this may happen, we illustrate them in
Fig.~\ref{fig:path} and in Fig.~\ref{fig:paths}. The first one occurs
when a path for the node pair is already included in another (previously
inserted) path into the layer.
For instance, after the path in the figure is inserted into layer $l$, all
sub-paths ($(v_2,v_4)$, $(v_3, v_4)$) become included as well, forcing $v_2$
and $v_3$ to route along minimal paths towards destination $v_4$ in layer $l$.

The second scenario occurs when no path of required length can be found because
routing via any of the source node's neighbors would result in a path too short
or too long. In our second example, the almost-minimal paths are constrained to
have length exactly $3$. At first, the two almost-minimal paths $q =
(v_1,v_2,v_3,u_3)$ and $q' = (w_1, w_2, w_3,u_3)$ are inserted, which fixes the
paths for all node pairs in the set $\{(v_i, u_3), (w_i, u_3) \mid i \in
\{1,2,3\}\}$. Now any path for the node pair $(u_1, u_3)$ that respects the
already inserted paths will have length $l \in \{1, 2, 4\}$ because it would
have to come from the following set of paths: $\{ (u_1, q)$, $(u_1, q')$,
$(u_1, u_3)$, $(u_1, u_2, u_3)$, $(u_1, u_2, v_2, v_3, u_3)$, $(u_1, u_2, w_2,
w_3, u_3) \}$. If this scenario occurs, we route minimally, i.e. path $(u_1, u_3)$.

\subsubsection{Specification of Forwarding Tables}
\
In layered routing, each forwarding entry
$(l,s,d) \in \text{layers} \times \text{switches} \times \text{switches}$
corresponds to the port that switch $s$ uses when routing in layer $l$ and
transmitting a packet addressed to a switch $d$.

\iftr
\else
\ifcr
\else
\subsection{Path Diversity vs.~Network Size}
\label{sec:max_net}
\
Increasing the number of different paths between each node pair
requires more layers and thus also more
addresses assigned to each node (i.e., a larger LMC value).
However, using more addresses within one node decreases the maximum number of
nodes that can be used in the network overall (because the address field size
is fixed to 16 bits).
We analyze this tradeoff in Table~\ref{tab:uff}. We assume the maximum SF network based
on 36-port switches, that guarantees full global bandwidth. The results
illustrate that one can use $4$ layers without having to make any compromises
on the networks size, but anything beyond $4$ layers would reduce the maximum
network size. At this point, the constraining factor is no longer the switch radix,
but the address space.
In Sections~\ref{theory} and~\ref{sec:eval}, we show that -- fortunately -- our
routing scheme's performance is already quite substantial with just 4 layers and does not need more than 8 layers for high performance.

\begin{table}[h]
\caption{\textmd{Maximum number of switches and servers supported by a
single-subnet SF-based IB network, while achieving full global bandwidth and
$2^{LMC}$ many addresses per switch. \textbf{``L''}: LMC, \textbf{``\#A''}:
Number of addresses per server.}} 
\centering
\setlength{\tabcolsep}{4pt}
\scriptsize
\begin{tabular}{llllllllllllll}
\toprule
    \multicolumn{2}{c}{} & \multicolumn{4}{c}{\textbf{36-port switches}} & \multicolumn{4}{c}{\textbf{48-port switches}} & \multicolumn{4}{c}{\textbf{64-port switches}} \\
    \textbf{L} & \textbf{\#A} & $N_r$ & $N$ & $k'$ & $p$ & $N_r$ & $N$ & $k'$ & $p$ & $N_r$ & $N$ & $k'$ & $p$\\ \midrule
    0 & 1 & 512 & 6144 & 24 & 12 & 882 & 14112 & 31 & 16 & 1568 & 32928 & 42 & 21\\
    1 & 2 & 512 & 6144 & 24 & 12 & 882 & 14112 & 31 & 16 & 1250 & 23750 & 37 & 19\\
    2 & 4 & 512 & 6144 & 24 & 12 & 800 & 12000 & 30 & 15 & 800 & 12000 & 30 & 15\\
    3 & 8 & 450 & 5400 & 23 & 12 & 450 & 5400 & 23 & 12 & 450 & 5400 & 23 & 12\\
    4 & 16 & 288 & 2592 & 18 & 9 & 288 & 2592 & 18 & 9 & 288 & 2592 & 18 & 9\\
    5 & 32 & 162 & 1134 & 13 & 7 & 162 & 1134 & 13 & 7 & 162 & 1134 & 13 & 7\\
    6 & 64 & 98 & 588 & 11 & 6 & 98 & 588 & 11 & 6 & 98 & 588 & 11 & 6\\
    7 & 128 & 72 & 360 & 9 & 5 & 72 & 360 & 9 & 5 & 72 & 360 & 9 & 5\\
\bottomrule
\end{tabular}%
\label{tab:uff}
\end{table}

\fi
\fi

\section{Additional Results}
\label{sec:add_res}

\subsection{Changes for Custom Alltoall}
\label{sec:a2a_improvement}

We decided not to use the OpenMPI's default implementation of alltoall, as the algorithms it relies on result in sub-optimal performance for the deployed SF. 
Empirically, we determined that the best-performing alltoall for our system was a simple algorithm that posts all non-blocking send and receive requests simultaneously and then waits for completion.
Other collectives did not show a similar impact, and we thus used the default implementations. These issues are not expected with newer hardware.

\begin{figure}[t]
\vspaceSQ{-1em}
\centering
\includegraphics[width=0.74\linewidth]{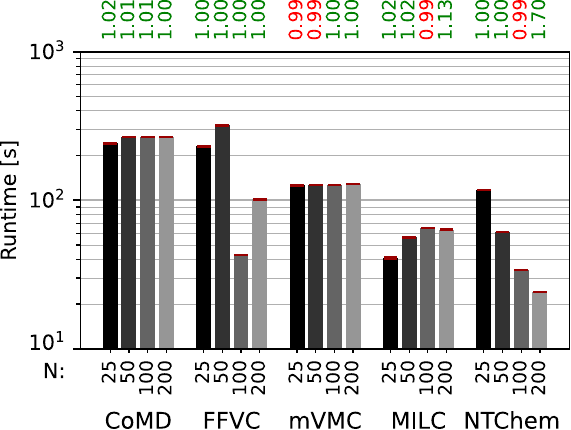}
\vspaceSQ{-1em}
\vspace{-0.5em}
\caption{\textmd{Runtime of scientific workloads (lower is better) - SF R vs. FT}}
\vspace{-1.1em}
\label{fig:sc_wor}
\end{figure}

\begin{figure}[t]
\vspaceSQ{-1em}
\centering
\vspaceSQ{0.3em}
\vspace{0.3em}
\begin{subfigure}[t]{0.46 \linewidth}
\captionsetup{justification=centering}
\centering
\includegraphics[width=\linewidth]{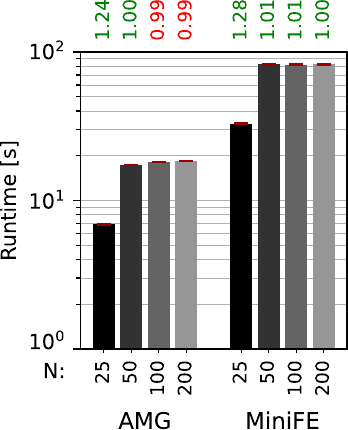}
\vspaceSQ{-1.0em}
\vspace{-1.0em}
\caption{\textmd{SF R vs. FT}}
\label{fig:sc_wor_rest_sf_r}
\end{subfigure}
\vspaceSQ{-1em}
\vspaceSQ{-1em}
\begin{subfigure}[t]{0.46 \linewidth}
\captionsetup{justification=centering}
\centering
\includegraphics[width=1.0\linewidth]{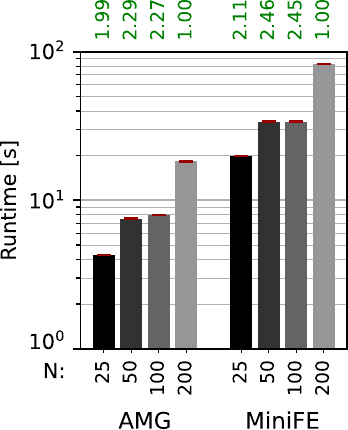}
\vspaceSQ{-1.0em}
\vspace{-1.0em}
\caption{\textmd{SF L vs. FT}}
\label{fig:sc_wor_rest_sf_l}
\end{subfigure}
\vspaceSQ{-0.5em}
\vspace{-0.5em}
\caption{\textmd{Runtime of additional scientific workloads (lower is better)}}
\vspace{-1em}
\vspaceSQ{-1em}
\label{fig:sc_wor_rest}
\end{figure}

\begin{figure}[t]
\vspaceSQ{-1em}
\centering
\includegraphics[width=0.64\linewidth]{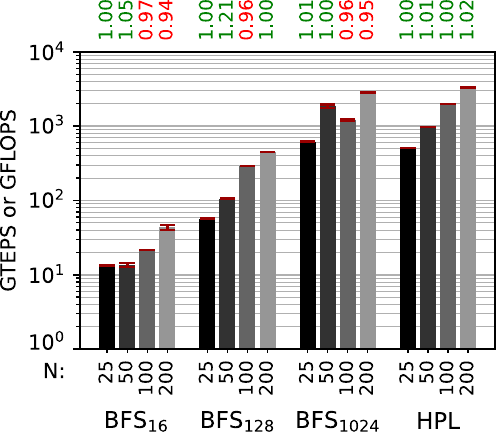}
\vspaceSQ{-0.5em}
\vspace{-0.5em}
\caption{\textmd{Performance of HPC benchmarks (higher is better) - SF R vs. FT}}
\vspace{-1.1em}
\label{fig:hpc_workloads}
\end{figure}

\begin{figure}[t]
\vspaceSQ{0.6em}
\vspace{0.6em}
\centering
\includegraphics[width=\linewidth]{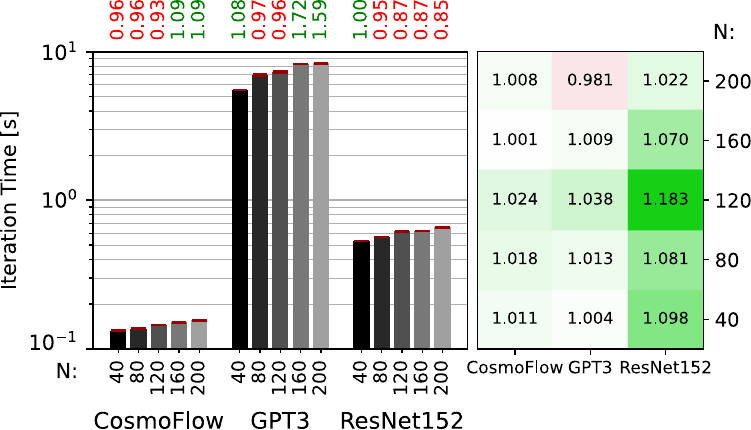}
\caption{\textmd{Iteration time of DNN proxy workloads (lower is better) SF R vs. FT and routing improvement of this work over DFSSSP (heatmap) in SF R}}
\label{fig:dl_combined_performance}
\vspace{-1.1em}
\end{figure}

\subsection{Scientific Workloads \& HPC Benchmarks}
\label{sec:add_res_sw_hpc}
We show in Fig.~\ref{fig:sc_wor} the runtime and relative performance of the solver/kernel for each of the scientific workloads on SF using the random placement strategy.
We observe similar trends as for the linear placement strategy for all scientific workloads and SF's performance aligns closely with FT's, while no significant speedup or slowdown through the use of non-minimal paths could be observed.

In Fig.\ref{fig:sc_wor_rest}, we present the relative performance of two additional scientific workloads, AMG\cite{HENSON2002155} and MiniFE~\cite{minife}, on SF, using both placement strategies. 
For this assessment, AMG was configured with a $128^3$ cube per process, while MiniFE was set with grid input dimensions of $n_{{x|y|z}_b} = 90$. 
In accordance with these configurations, clear weak-scaling behavior is evident under the random placement strategy.
On the other hand, with the linear placement strategy, the observed trends are less distinct, and, there are instances where SF outperformed FT by unexpected margins.
We believe that this disparity can't be merely attributed to the variations in communication stemming from the placement strategy, as the applications in consideration aren't generally communication-bound and the FT is fully non-blocking. However, the precise cause remains unclear.

Fig.~\ref{fig:hpc_workloads} shows the performance of the HPC benchmarks on SF using the random placement strategy, 
results that largely mirror those obtained using the linear placement strategy.

\subsection{Deep Learning Workloads}

The left part of Fig.~\ref{fig:dl_combined_performance} shows the runtime and relative performance of the DNN proxies with the random placement strategy. The results are also very similar to those obtained using the linear placement strategy, including GPT-3 matching the performance trends of the MPI Allreduce pattern with the random placement strategy and comparable node counfigurations (cf. Fig.~\ref{fig:speedup_imb_allr_sf_r}).

However, similar to previous results, the right part of Fig.~\ref{fig:dl_combined_performance} shows that our work generally matches or outperforms DFSSSP, achieving up to a 1.18x speedup. 

\section{Pricing details}
\label{sec:costs}

We based our pricing on data colfaxdirect.com\footnote{\href{http://www.colfaxdirect.com}{COLFAX DIRECT website}} and SHI.com\footnote{\href{https://www.shi.com/}{SHI website}}. Regarding the equipment selection, we use InfiniBand Topology Configurator \footnote{\href{https://www.nvidia.com/en-us/networking/infiniband-configurator/}{Mellanox InfiniBand Topology Generator tool}}. For different switch sizes, we selected different models from current Nvidia offerings. For example, for a 36-port switch, we chose Mellanox SB7800 EDR 100Gb/s\footnote{\href{http://www.colfaxdirect.com/store/pc/viewPrd.asp?idproduct=3049}{Mellanox SB7800 EDR
100Gb/s product detail}}. For a 40-port switch, we decided to use Mellanox Quantum QM8700 HDR 200Gb/s \footnote{\href{http://www.colfaxdirect.com/store/pc/viewPrd.asp?idproduct=3685}{Mellanox
Quantum QM8700 HDR 200Gb/s product detail}}. Finally, for a 64-port switch, we use Nvidia QM9700 NDR 400G \footnote{\href{http://www.colfaxdirect.com/store/pc/viewPrd.asp?idproduct=4165&idcategory=0}{ Nvidia QM9700 NDR 400G product detail}} model. For AoC cables, we selected active fiber links, and for DAC cables, we chose passive copper cables for endpoint connections. Again, we base our estimations on mentioned earlier InfiniBand Topology Configurator online service. However, it can be challenging to determine the cost of networking hardware because the prices of such hardware can vary greatly depending on the quantity ordered, and large orders may be eligible for substantial discounts.

\end{document}